\documentclass[%
 preprint, 
superscriptaddress,
 aps, prx, longbibliography
]{revtex4-2}

\usepackage{graphicx}
\usepackage{dcolumn}
\usepackage{bm}
\usepackage{graphicx} 
\usepackage{amsmath}
\usepackage{upgreek}
\usepackage{amsfonts}
\usepackage{amssymb}
\usepackage[dvipsnames]{xcolor}

\newcommand{\ket}[1]{\lvert #1 \rangle}
\newcommand{\bra}[1]{\langle  #1 \lvert}

\begin{document}

\preprint{APS/123-QED}

\title{\textbf{Benchmarking Gaussian and non‑Gaussian input states with a hybrid sampling platform} 
}%

\author{Michael Stefszky}
\altaffiliation{These authors contributed equally to this work}
\affiliation{Paderborn University, Integrated Quantum Optics, Warburger Str. 100, 33098 Paderborn, Germany}
\affiliation{Paderborn University, Institute for Photonic Quantum Systems (PhoQS), Warburger Str. 100, 33098 Paderborn, Germany}
 
\author{Kai-Hong Luo}
\altaffiliation{These authors contributed equally to this work}
\affiliation{Paderborn University, Integrated Quantum Optics, Warburger Str. 100, 33098 Paderborn, Germany}
\affiliation{Paderborn University, Institute for Photonic Quantum Systems (PhoQS), Warburger Str. 100, 33098 Paderborn, Germany}

\author{Jan-Lucas Eickmann}
\altaffiliation{These authors contributed equally to this work}
\affiliation{Paderborn University, Integrated Quantum Optics, Warburger Str. 100, 33098 Paderborn, Germany}
\affiliation{Paderborn University, Institute for Photonic Quantum Systems (PhoQS), Warburger Str. 100, 33098 Paderborn, Germany}

\author{Simone Atzeni}
\altaffiliation{These authors contributed equally to this work}
\affiliation{Paderborn University, Integrated Quantum Optics, Warburger Str. 100, 33098 Paderborn, Germany}
\affiliation{Paderborn University, Institute for Photonic Quantum Systems (PhoQS), Warburger Str. 100, 33098 Paderborn, Germany}

\author{Florian Lütkewitte}
\affiliation{Paderborn University, Integrated Quantum Optics, Warburger Str. 100, 33098 Paderborn, Germany}
\affiliation{Paderborn University, Institute for Photonic Quantum Systems (PhoQS), Warburger Str. 100, 33098 Paderborn, Germany}

\author{Cheeranjiv Pandey}
\affiliation{Paderborn University, Integrated Quantum Optics, Warburger Str. 100, 33098 Paderborn, Germany}
\affiliation{Paderborn University, Institute for Photonic Quantum Systems (PhoQS), Warburger Str. 100, 33098 Paderborn, Germany}

\author{Fabian Schlue}
\affiliation{Paderborn University, Integrated Quantum Optics, Warburger Str. 100, 33098 Paderborn, Germany}
\affiliation{Paderborn University, Institute for Photonic Quantum Systems (PhoQS), Warburger Str. 100, 33098 Paderborn, Germany}

\author{Jonas Lammers}
\affiliation{Paderborn University, Integrated Quantum Optics, Warburger Str. 100, 33098 Paderborn, Germany}
\affiliation{Paderborn University, Institute for Photonic Quantum Systems (PhoQS), Warburger Str. 100, 33098 Paderborn, Germany}

\author{Mikhail Roiz}
\affiliation{Paderborn University, Integrated Quantum Optics, Warburger Str. 100, 33098 Paderborn, Germany}
\affiliation{Paderborn University, Institute for Photonic Quantum Systems (PhoQS), Warburger Str. 100, 33098 Paderborn, Germany}

\author{Timon Schapeler}
\affiliation{Paderborn University, Department of Physics, Warburger Str. 100, 33098 Paderborn, Germany}
\affiliation{Paderborn University, Institute for Photonic Quantum Systems (PhoQS), Warburger Str. 100, 33098 Paderborn, Germany}

\author{Laura Ares}
\affiliation{Paderborn University, Theoretical Quantum Science, Warburger Str. 100, 33098 Paderborn, Germany}
\affiliation{Paderborn University, Institute for Photonic Quantum Systems (PhoQS), Warburger Str. 100, 33098 Paderborn, Germany}

\author{Milad Yahyapour}
\affiliation{Menlo Systems GmbH, Bunsenstr. 5, 82152 Planegg, Germany}

\author{Alexander Kastner}
\affiliation{Menlo Systems GmbH, Bunsenstr. 5, 82152 Planegg, Germany}

\author{Joschua Martinek}
\affiliation{Menlo Systems GmbH, Bunsenstr. 5, 82152 Planegg, Germany}

\author{Michael Mittermair}
\affiliation{Menlo Systems GmbH, Bunsenstr. 5, 82152 Planegg, Germany}

\author{Carlos Sevilla-Guti\'{e}rrez}
\affiliation{Institute of Applied Physics, Friedrich Schiller University Jena, Germany}
\affiliation{Fraunhofer Institute for Applied Optics and Precision Engineering IOF, Germany}

\author{Marius Leyendecker}
\affiliation{Institute of Applied Physics, Friedrich Schiller University Jena, Germany}
\affiliation{Fraunhofer Institute for Applied Optics and Precision Engineering IOF, Germany}

\author{Oskar Kohout}
\affiliation{Institute of Applied Physics, Friedrich Schiller University Jena, Germany}
\affiliation{Fraunhofer Institute for Applied Optics and Precision Engineering IOF, Germany}

\author{Dmitriy Mitin}
\affiliation{Institute of Applied Physics, Friedrich Schiller University Jena, Germany}
\affiliation{Fraunhofer Institute for Applied Optics and Precision Engineering IOF, Germany}

\author{Ronald Holzwarth}
\affiliation{Menlo Systems GmbH, Bunsenstr. 5, 82152 Planegg, Germany}

\author{Jan Sperling}
\affiliation{Paderborn University, Theoretical Quantum Science, Warburger Str. 100, 33098 Paderborn, Germany}
\affiliation{Paderborn University, Institute for Photonic Quantum Systems (PhoQS), Warburger Str. 100, 33098 Paderborn, Germany}

\author{Tim Bartley}
\affiliation{Paderborn University, Department of Physics, Warburger Str. 100, 33098 Paderborn, Germany}
\affiliation{Paderborn University, Institute for Photonic Quantum Systems (PhoQS), Warburger Str. 100, 33098 Paderborn, Germany}

\author{Fabian Steinlechner}
\affiliation{Institute of Applied Physics, Friedrich Schiller University Jena, Germany}
\affiliation{Fraunhofer Institute for Applied Optics and Precision Engineering IOF, Germany}

\author{Benjamin Brecht}
\affiliation{Paderborn University, Integrated Quantum Optics, Warburger Str. 100, 33098 Paderborn, Germany}
\affiliation{Paderborn University, Institute for Photonic Quantum Systems (PhoQS), Warburger Str. 100, 33098 Paderborn, Germany}

\author{Christine Silberhorn}
\affiliation{Paderborn University, Integrated Quantum Optics, Warburger Str. 100, 33098 Paderborn, Germany}
\affiliation{Paderborn University, Institute for Photonic Quantum Systems (PhoQS), Warburger Str. 100, 33098 Paderborn, Germany}



\date{\today}

\begin{abstract}
The original boson sampling paradigm---consisting of multiple single-photon input states, a large interferometer, and multi‑channel click detection---was originally proposed as a photonic route to quantum computational advantage. Its non‑Gaussian resources, essential for outperforming any classical system, are provided by single‑photon inputs and click detection. Yet the drive toward larger experiments has led to the replacement of experimentally demanding single‑photon sources with Gaussian states, thereby diminishing the available non‑Gaussianity---a critical quantum resource. As the community broadens its focus from the initial sampling task to possible real‑world applications, it becomes crucial to quantify the performance cost associated with reducing non‑Gaussian resources and to benchmark sampling platforms that employ different input states.
To address this need, we introduce the Paderborn Quantum Sampler (PaQS), a hybrid platform capable of performing sampling experiments with eight Gaussian or non‑Gaussian input states in a 12‑mode interferometer within a single experimental run. This architecture enables direct, side‑by‑side benchmarking of distinct sampling regimes under otherwise identical conditions -- only the input states vary between configurations. By employing a semi-device-independent framework based on matrix of moments approaches that offers certification that does not rely on prior knowledge of the interferometer or the input states, we verify that the observed data cannot be reproduced using classical resource states—a prerequisite for demonstrating quantum advantage. Applying this framework, we show that non-Gaussian inputs offer a distinct advantage in the verification of nonclassicality under our proposed metric.
\end{abstract}

\maketitle


\section{Introduction}

Boson sampling (BS), now often termed Fock‑state boson sampling, was first proposed as a route to demonstrating a provable quantum computational advantage \cite{Aaronson2010,Aaronson2013}. Early proof‑of‑concept experiments quickly followed, but their scale was limited by the difficulty involved in generating many high‑quality single‑photon inputs \cite{Broome2013,Spring2013,Tillmann2013,Spagnolo2014}. Researchers began to explore the use of alternative input states—such as two-mode squeezed-vacuum (TMSV) states in the protocol now known as scattershot boson sampling (SBS)—as a means of reducing the experimental overhead \cite{Lund2014}. Ultimately, the issue of system scaling was resolved when it was revealed that driving the system with single-mode squeezed-vacuum (SMSV) states preserves the computational complexity of the original proposal without the need for non-Gaussian input resources \cite{Hamilton2017,Kruse2019}. The resulting protocol, Gaussian boson sampling (GBS), enabled a sequence of increasingly large experiments that pushed steadily toward a robust demonstration of quantum computational advantage \cite{Paesani2019,Zhong2021,Madsen2022,Deng2023,Yu2023,Zhu2024,Liu2026}.

Concurrently, researchers began investigating the utility of sampling architectures for addressing tasks with real‑world relevance. Notably, problems such as molecular vibronic spectroscopy \cite{Huh2015} and a variety of graph‑theoretic computations were mapped onto GBS-type devices \cite{Banchi2020,Arrazola2021,Bradler2021,Deng2023Graph,Anteneh2023}, and dedicated implementations successfully solved these tasks \cite{Yu2023,Zhu2024}. It remained unclear, however, whether these problems could exhibit a genuine quantum computational advantage in this setting. Subsequent work revealed that many could, in fact, be solved efficiently using classical means \cite{Oh2021,Oh2024,Oh2024Alg,Lim2025,Cilluffo2026}. This realization has led to two key outcomes: it has motivated investigations into alternative applications such as Monte Carlo integration \cite{Andersen2025,Andersen2025est,Anguita2025}, quantum field dynamics \cite{Dachille2026}, and machine learning \cite{Nokkala2021,Hoch2025,Cimini2025,Sakurai2025}, and it has stimulated deeper exploration of integrating additional non-Gaussian resources in these system to broaden their computational capabilities. Furthermore, techniques for overcoming system imperfections, such as loss, are being developed \cite{Ellenberg26}.

The non‑Gaussian element in GBS systems arises from photon‑number‑resolved (PNR) detection (or in earlier implementations, click detection). Extending the computational capabilities of such systems is therefore most easily achieved by supplementing this resource, either through advanced operations within the interferometer \cite{Hoch2025,Barkhofen2017,Spagnolo2023,Monbroussou2025}, or by driving the interferometer with more exotic non‑Gaussian input states, whose impact has been the subject of extensive investigation \cite{Rohde2015,Chabaud2020,Chabaud2021,Chabaud2021Sim,Chabaud2023,Crescimanna2024,Hamilton2025,Bianchi2025,Oh2025}. Of course, while these advanced operations and resource states increase the capabilities of the system, they generally come at the cost of reduced scalability. Consequently, the next generation of sampling devices will require a conceptual redesign of both their system architecture and resource allocation.

A corresponding rethink is necessary for benchmarking the performance of these newly conceived sampling systems. A commonly used verification technique is to quantify the closeness of generated data to that which is expected from theory, typically implemented using a metric known as the total variation distance \cite{Aaronson2010}.Although this metric can be used to make strong statements about computational complexity, this method requires exponentially many samples and, therefore, cannot be used with large systems. More advanced techniques utilizing phase-space representations can be used to overcome this scaling issue \cite{Dellios2025,Dellios2025_QST}. However, the aforementioned validation techniques are sensitive to experimental imperfections; errors in implementing the multi-mode interferometer are critical.
Progressing the field therefore requires new certification tools capable of identifying genuinely non‑classical features---or quantumness---and experimental platforms that enable fair, side‑by‑side comparison of different architectures. These measures should provide some degree of insensitivity to experimental imperfections and should be accessible from the generated data.

\begin{figure}[!ht]
    \centering
    \includegraphics[width = 0.8\textwidth]{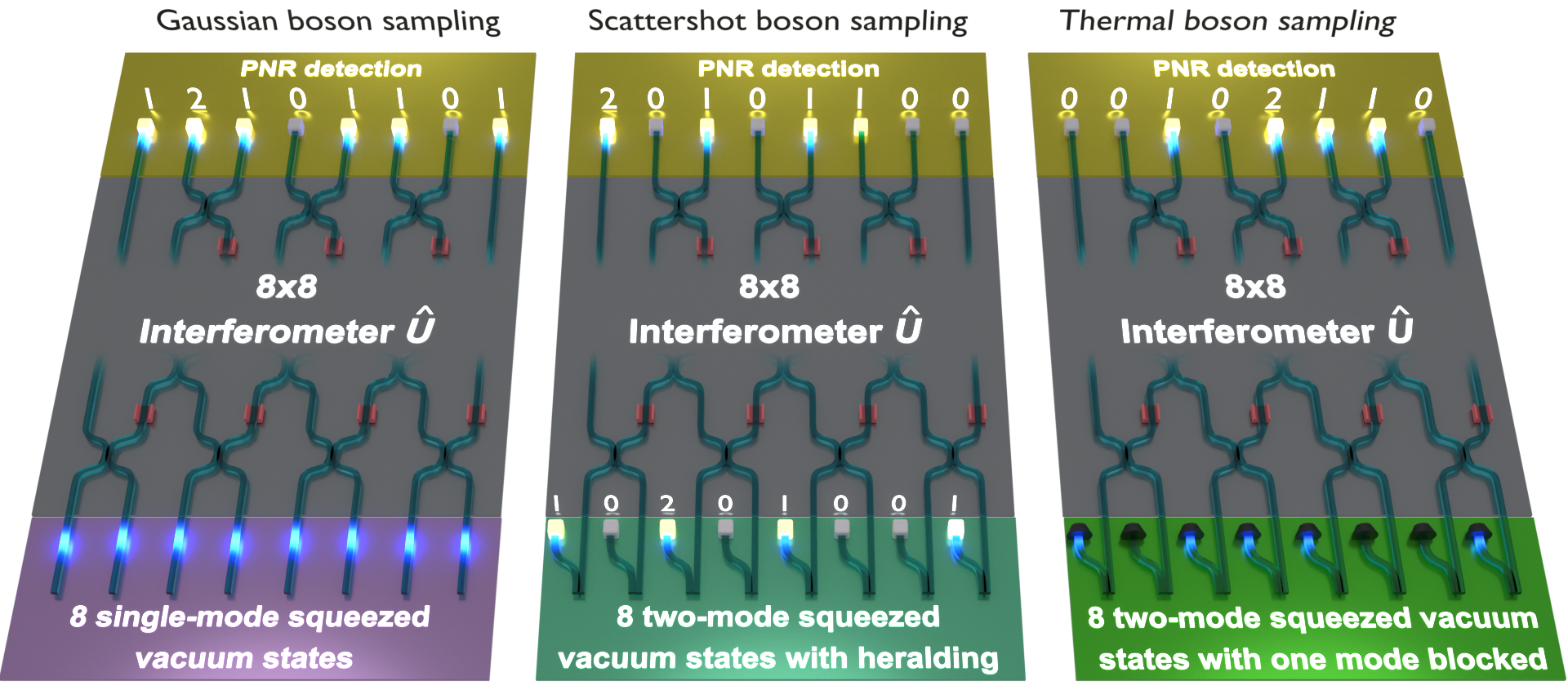}
    \caption{Conceptual implementations for various sampling configurations.
    In all cases, the input fields enter from the bottom of the setup and are detected with PNR detectors upon exiting the interferometer at the top.
    The input states of the interferometer can be either eight SMSV states, heralded Fock states, or thermal states, corresponding to implementing GBS, SBS, and thermal boson sampling (TBS), respectively. 
    To implement SBS, the second mode of each TMSV input state is detected with PNR detectors to herald the number of photons entering each interferometer port. 
    To realize sampling with thermal states the second mode of each TMSV state is discarded, i.e., traced out.}
    \label{Fig:GBSSBSconcept}
\end{figure}

In this work, we introduce the \textit{Paderborn Quantum Sampler} (PaQS), an experimental platform designed to benchmark multiple sampling schemes within a single experimental run. To compare the performance associated with different input‑state resources---both Gaussian and non-Gaussian---as illustrated in Figure \ref{Fig:GBSSBSconcept}, we further develop a benchmarking framework based on normally ordered moments of the photon‑number operator \cite{Walschaers2016,SRV05, Agarwal1992, Shchesnovich2016, Phillips2019,Vandermeer2021,Heinzel26}. This framework, in contrast to previous approaches that attempt to directly certify computational complexity, instead certifies the presence of quantumness in generated data -- a necessary, but not sufficient condition for quantum computational complexity ~\cite{Rahimi2016}. The advantage of this approach is that it is robust to experimental imperfections (semi-device-independent as knowledge of the implemented unitary transformation and the input states is not required) and can be directly applied to data generated from the boson sampling platform — unlike measures such as Wigner function negativity \cite{Mari2012}, which require additional, complex measurement apparatus.

Although one might expect,\textit{ a priori}, that Fock states would outperform squeezed states due to their higher degree of nonclassicality \cite{Lee1994}, the situation is more nuanced: it has been shown that the specific type of nonclassicality that can be exploited for a given task is not necessarily correlated with the ``absolute'' amount of nonclassicality \cite{Sperling2019,Aasi2013}. Therefore, understanding the benefits offered by distinct input states is critical: in fact, our results reveal pronounced differences between the generated GBS and SBS datasets. The quantumness of SBS data is seen to steadily increase with the mean photon number of the input states, whereas the GBS data displays strong non‑classical signatures at low mean photon numbers, but fails to maintain them as the brightness increases -- even though it is known that the nonclassicality of squeezed states increases with increasing degree of squeezing \cite{Lee1994}. This divergence underscores the fundamentally different behaviors arising from distinct input‑state resources and highlights the need for a conceptual shift when evaluating the performance of boson sampling architectures.

\section{Experimental Setup}
\begin{figure}[!ht]
    \centering
    \includegraphics[width = 0.9\textwidth]{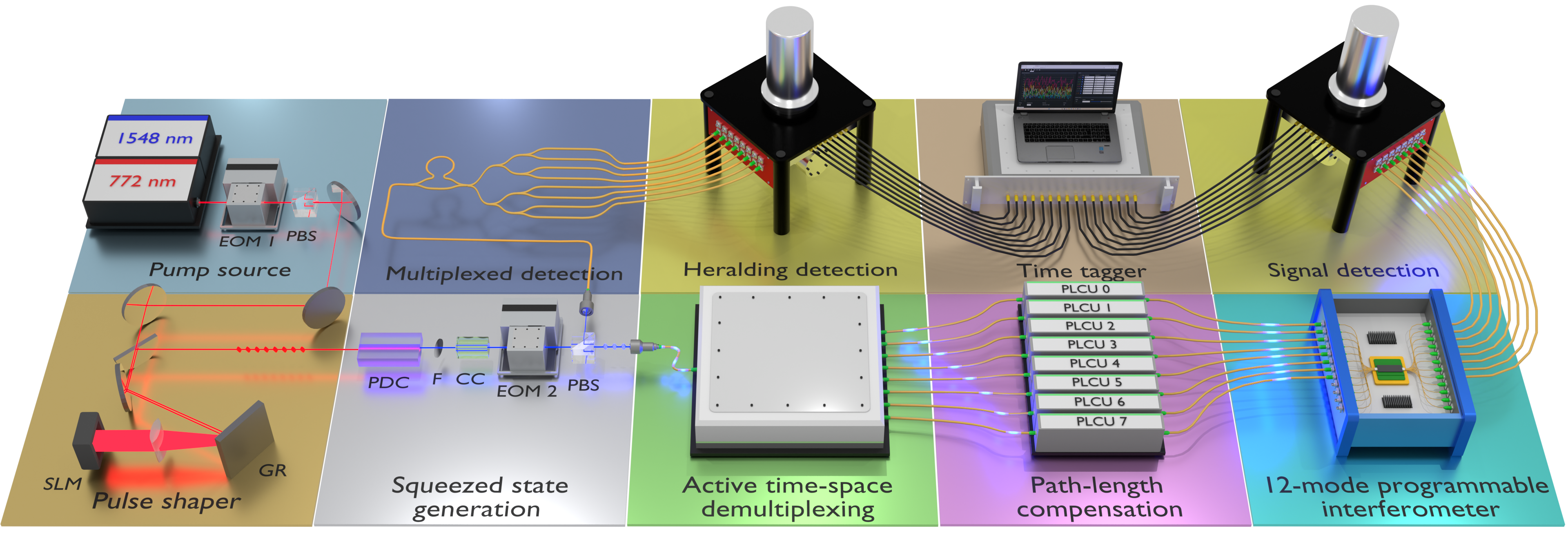}
    \caption{Schematic of the PaQS sampling system. 
    The subsystems are separated to highlight the modular design. 
    EOM — electro-optic modulator; PBS — polarizing beam splitter; SLM — spatial light modulator; GR — grating; PDC — parametric down-conversion source; F — filtering stage; CC — (temporal) compensation crystal; PLCU — path-length compensation unit.}
    \label{Fig:GBSsetup}
\end{figure}

Figure~\ref{Fig:GBSsetup} presents a schematic of the PaQS system, divided into its constituent subsystems for clarity. The design of PaQS emphasizes integration: both the squeezed-light source (or parametric down-conversion source, PDC) and the programmable interferometer are realized on integrated photonic platforms. Notably, the integrated source, in conjunction with an electro-optic modulator and polarizing beamsplitter, enables dynamic switching between TMSV- and SMSV-state generation. This architecture allows multiple sampling configurations to be acquired during a single experimental run, thereby ensuring that results from the different sampling schemes are directly commensurable.
Furthermore, the picosecond squeezed-light pulses generated by the waveguided source enables the implementation of intrinsic photon-number resolution (PNR) in the employed superconducting nanowire single-photon detectors (SNSPDs). The integrated interferometer offers convenient and complete programmability of the implemented unitary transformation, with an average insertion loss below 3\,dB. An overview of the key subsystems of PaQS is given in Figure~\ref{Fig:GBSsetup} and described here, while detailed descriptions can be found in the Appendixes.

\subsection{Pump Source}

Pump pulses are generated by a newly developed laser system that begins with a SmartComb (Menlo Systems GmbH) system producing femtosecond pulses in the telecom C‑band. Light from the SmartComb enters a laser extension unit that provides optical gain and generates various control signals. The amplified pulses are then frequency‑doubled in a free‑space second‑harmonic stage to yield picosecond pulses centered around 772\,nm with a full-width at half maximum bandwidth of approximately 1\,nm, a repetition rate of 80\,MHz and pulse energies up to approximately 12\,nJ. 

The pulses are passed through a pulse‑picker (EOM1, QUBIG GmbH HVOS‑NIR‑1.5k100) that selects any desired number of pulses from the 80\,MHz train every 1\,$\upmu$s. 
The number of selected pulses defines the number of squeezed‑light pulses per experimental cycle, while the pulse‑train generation rate sets the overall system sampling rate of 1\,MHz. 
The pump field is then spectrally shaped in a folded 4f‑line pulse shaper composed of a grating (groove spacing $1/2200$\,mm) and a programmable spatial light modulator (Santec SLM200) to ensure single spectro-temporal-mode operation of the generated squeezed light~\cite{Christ2011}.

\subsection{Squeezed-State Generation}

The prepared pump‑pulse train enters the 20\,mm‑long waveguided PPKTP crystal (AdvR Inc.) designed for a type‑\MakeUppercase{\romannumeral 2} PDC process that creates spectrally indistinguishable, ps-long signal and idler fields in a single spectro-temporal mode. After the waveguide crystal, the output fields pass through a filtering stage composed of a silicon window that suppresses the pump and a 2\,nm spectral filter that removes phase matching side lobes. 
An 8.8\,mm‑long KTP group‑delay compensation crystal (CC) then corrects for birefringent walk‑off accumulated in the waveguide, ensuring temporal overlap of the signal and idler pulses. 
The filtered, temporally synchronized fields subsequently enter an electro‑optic modulator (EOM2, QUBIG GmbH HVOS‑SWIR‑1.5k100), which sets the output polarization of the transmitted modes on a polarizing beam splitter (PBS). By varying the voltage applied to EOM2, we can rapidly switch (rise/fall time of 4\,ns) between generating two independent SMSV states and a TMSV state at the PBS output.
Interference of the two polarized fields at the PBS produces independent SMSV states used for GBS implementations~\cite{Hamilton2017}, whereas deterministic separation of the fields yields a TMSV state used to realize SBS~\cite{Lund2014} or, by using the same data and disregarding the heralding information, TBS~\cite{Keshari2015} configurations in a single experimental data run. The fields at both PBS output ports are fiber‑coupled. 


\subsection{Multiplexed Detector and Heralding}

A multiplexed detection module attached to one PBS output both spatially and temporally separates the incoming pulses to measure the photon number in this heralding arm. 
This module enables photon‑number heralding for implementing higher-order SBS and allows verification of the lack of correlations between the two PBS outputs in the GBS configuration. 
Discarding the heralding component of the data obtained in SBS effectively prepares thermal input states for the interferometer~\cite{Christ2011}. 

The multiplexed architecture is required because, in the absence of a second time‑to‑space demultiplexer, the 12\,ns pulse separation is shorter than the SNSPD dead time of approximately 80\,ns. 
The implemented scheme minimizes the probability of photon arrival during detector dead time, achieving an average heralding efficiency of $38.4\,\%\!\pm\!0.4\,\%$.

\subsection{Time‑to‑Space Demultiplexing}
The fiber-coupled light that exited the second PBS output passes through an active time‑to‑space demultiplexer (DMX16\textunderscore80, QUBIG~GmbH), which routes each input pulse into individual output fibers via an EOM switching tree with an efficiency exceeding 80\%. 
Each output fiber is fed into a temperature‑stabilized path‑length compensation unit (PLCU, Menlo Systems GmbH) consisting of segments of fiber of variable length. 
These include approximately 60\,m of fiber wound around a piezo actuator and additional lengths compensating the 12.5\,ns time difference between consecutive pulses. 
The optical path length can be tuned using temperature control and mechanical strain. 

The phase of the fields exiting the PLCUs is currently free‑running; no active phase stabilization is implemented. 
Nonetheless, each PLCU can serve as an actuator for future active control once an appropriate error signal is defined. 
For the present implementation, all phase‑sensitive components are enclosed in polyethylene foam housing to achieve high passive stability.

\subsection{Integrated Interferometer}
The temporally synchronized pulses emerging from the demultiplexer are fed via separate fibers into a 12‑mode integrated interferometer (QUIX Quantum) that provides full programmability of the implemented unitary transformation. This interferometer is realized in the low-loss silicon nitride platform and exhibits an average insertion loss of $2.87\!\pm\!0.37$\,dB. 
The mean similarity between implemented and target unitary matrices drawn from a Haar‑random ensemble is approximately 94\% for 12‑dimensional and 97\% for 8‑dimensional transformations.


\subsection{Photon‑Number Detection}
Photon numbers in the eight utilized output modes of the interferometer are measured using eight SNSPDs (Single Quantum) with quantum efficiency exceeding 90\% and timing jitter below 20\,ps. 
An intrinsic PNR scheme is implemented, relying on the $\approx2$\,ps pulse duration of the squeezed‑light states and the sub‑2\,ps jitter of the Swabian Instruments Time Tagger X~\cite{Sauer2023,Schapeler2024}. 
Using this technique, our implementation discriminates events with up to three photons per pulse with high confidence~\cite{Schapeler2024,Sidorova2025}.


\section{System Characterization}
We summarize below the principal methods used to characterize and verify the operation of the PaQS subsystems. The corresponding results are shown in Figure~\ref{Fig:PaQSChar}, with further details provided in Appendix \ref{App:expdesign}.

The transmission of PaQS is quantified using the Klyshko method~\cite{Klyshko1980}, applied individually to each signal mode. This technique provides the total transmission of each signal mode from state generation until detection. The system is configured for SBS operation in the low-gain regime, and an identity transformation is programmed on the interferometer. The Klyshko efficiency $\eta^{(i)} = C_i/H_i$ for each mode~$i$ is given by the ratio of the measured coincidence rate $C_i$, between the relevant heralding bins and the output detector for that mode, and the single-detection rate $H_i$, in the heralding arm. The results, presented in Figure~\ref{Fig:PaQSChar}\,a, show that all modes achieve efficiencies above $6.5\%$, with an average of $8.7\%\!\pm\!1.5\%$.

Single spectro-temporal-mode operation of our type-\MakeUppercase{\romannumeral 2} waveguide source is verified through second-order autocorrelation measurements~\cite{Christ2011}. The system operates in SBS mode while photon statistics are recorded for one output of the PBS as the mean photon number of the generated state is varied by adjusting the pump power driving the nonlinear process. The generated mean photon number is determined by accounting for the loss on the measured state using the method described in Appendix \ref{App:measurement}. As shown in Figure~\ref{Fig:PaQSChar}\,b, the second-order correlation function converges to $g^{(2)} = 1.95 \pm 0.03$ at higher power, corresponding to an expected Schmidt mode number $K = 1.05 \pm 0.03$ effective modes—confirming both spectro-temporal purity and suitability for sampling experiments. The increase in $g^{(2)}$ at low mean photon numbers originates from a residual SMSV contribution because of imperfect polarization separation at the PBS (see Appendix \ref{App:expdesign}).

Next, we assess the ability of PaQS to switch between SMSV and TMSV generation. Operating in the low‑gain regime, we record the coincidence counts $n_\mathrm{C}$ between the herald and signal detectors as the driving voltage of EOM2’s high-voltage amplifier is varied (Figure~\ref{Fig:PaQSChar}\,c). The normalized coincidence signal varies nearly sinusoidally with a fitted visibility of $V = 96.3\%\!\pm\!1.2\%$ (See Appendix \ref{App:expdesign}), demonstrating precise and continuous tunability between SBS (maximum coincidences) and GBS (minimum coincidences) configurations. The difference in number of normalized coincidence counts for the two displayed SBS configurations (at 0 V and $\approx$1.35 V) is due to the different spatial-mode profiles of the two polarization modes within the waveguide.

Finally, Figure~\ref{Fig:PaQSChar}\,d illustrates the precision with which we can tune the optical path-length difference between input modes using the PLCUs in order to optimize the indistinguishability between different sources (further details in Appendix \ref{App:expdesign}). 
Implementing the SBS configuration in the low‑gain regime, we set the interferometer to realize a 50{:}50 beam splitter between any pair of chosen input modes. By varying the temperature of one PLCU fiber, we tune its optical length and record the resulting four‑fold coincidence counts $n_\mathrm{Corr}$ between the two beam-splitter outputs and their corresponding heralds, correcting for higher photon‑number contributions as described in Appendix \ref{App:expdesign}. 
This configuration effectively realizes Hong–Ou–Mandel (HOM) interference between distinct sources~\cite{ou1999photon, mosley2008conditional, jin2015spectrally}. 
Given the measured spectro-temporal mode number $K = 1.05$ from Figure~\ref{Fig:PaQSChar}\,b, the expected maximum visibility is $V \approx 95\%$. Some measurements approach this limit, while slight reductions in others are attributed primarily to imperfections in specific 50{:}50 beam splitters implemented in the interferometer.

\begin{figure}[!ht]
    \centering
    \includegraphics[width = 0.8\textwidth]{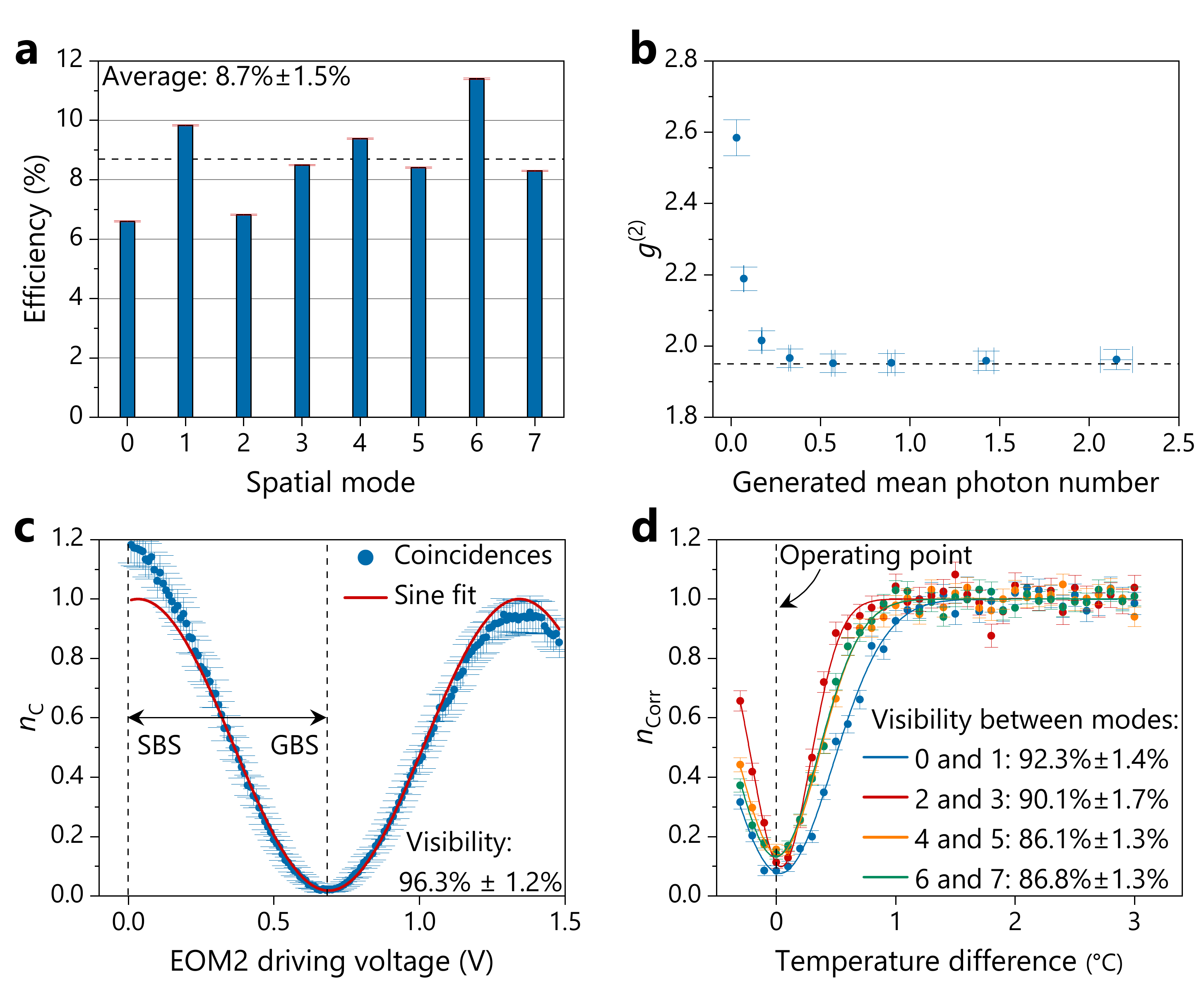}
    \caption{System verification measurements.
    \textbf{a,} Measured Klyshko efficiencies for each interferometer output mode. Average value of 8.7\%$\pm$1.5\% is indicated by the dashed line
    \textbf{b,} Measured second‑order correlation function from one arm of a TMSV state as the mean photon number of the state generated in the waveguide is varied. The asymptote at approximately 1.95 is indicated by the dashed line.
    \textbf{c,} Normalized coincidence counts between the heralding and signal detections as the EOM2 driving voltage is tuned. 
    The minima correspond to GBS and the maxima to SBS. 
    \textbf{d,} Coincidence counts (corrected for multiphoton events and normalized to the sinusoidal fit) between selected output detectors as the fiber temperature of one mode is varied, demonstrating HOM interference. 
    All uncertainties arise from Poisson counting statistics.}
    \label{Fig:PaQSChar}
\end{figure}

\section{Violation of Classicality Bounds}
We now apply the developed benchmarking framework to experimental data produced by PaQS, which directly probes for the presence of nonclassical photon‑number correlations between output modes, or quantumness. This approach leverages the fact that quantumness is a prerequisite for any genuine quantum computation or for generating results that cannot be efficiently simulated~\cite{Feynman1982}. In this way, the verification of nonclassicality is a necessary but not sufficient condition for quantum computational complexity. The proposed bound is independent of the implemented unitary transformation, making the analysis semi-device-independent and therefore robust against certain experimental imperfections. Moreover, the ability of PaQS to switch rapidly between GBS and SBS/TBS enables a fair benchmarking between these sampling regimes.

Our nonclassicality criterion is grounded in Glauber’s coherence theory~\cite{Glauber1963}. Here, we provide a brief description, while the complete derivation can be found in Appendix \ref{App:theory}. Recasting matrix of moments approaches~\cite{SRV05,Agarwal1992}, we identify quantumness in the photon‑number covariances $\mathrm{Cov}(n_j, n_k) = \langle \hat n_j \hat n_k \rangle - \langle \hat n_j \rangle \langle \hat n_k \rangle$ between output modes $j,k$. For classical light, the $M$‑mode matrix $(\mathrm{Cov}(n_j,n_k) - \delta_{j,k}\langle \hat n_j \rangle)_{j,k=1,\ldots,M}$, where $\delta_{j,k}$ denotes the Kronecker delta, is positive semidefinite. Therefore, observation of a negative eigenvalue constitutes unambiguous evidence of nonclassical photon‑number correlations and, therefore, the presence of quantumness in the system.

\begin{figure}[!ht]
    \centering
    \includegraphics[width = 1\textwidth]{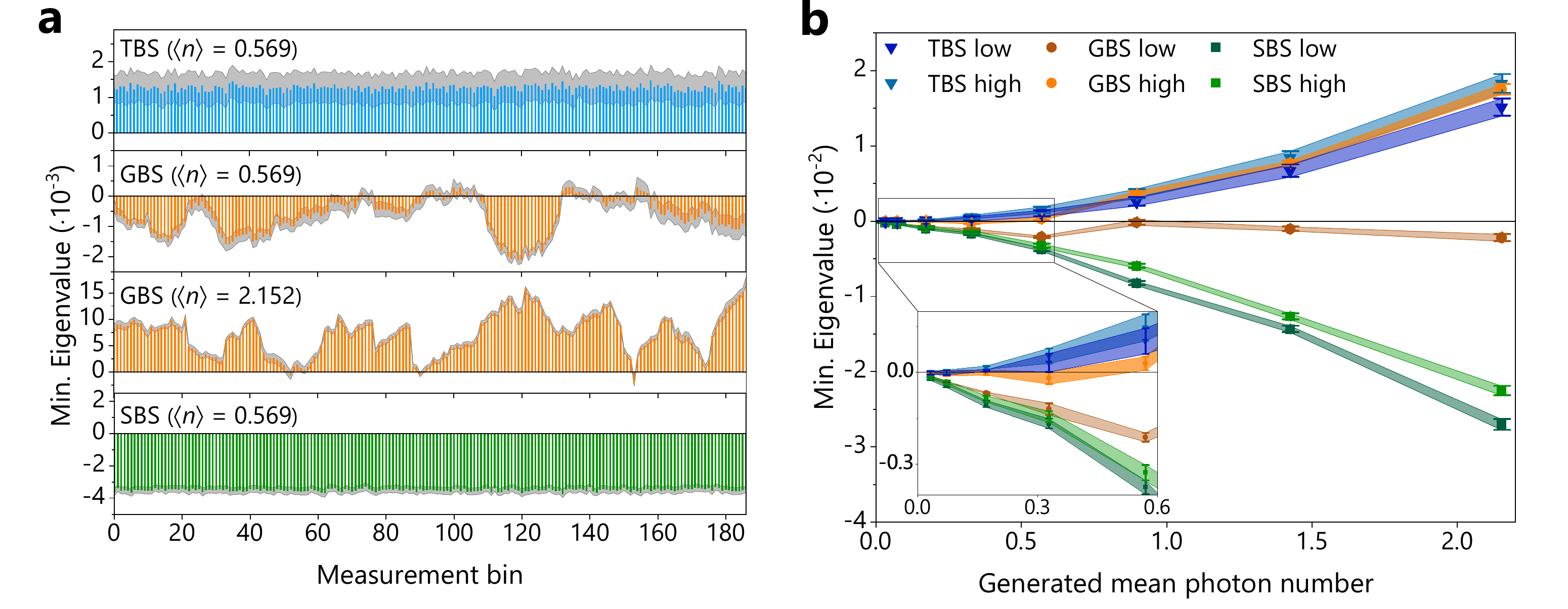}
    \caption{Quantumness analysis.
    \textbf{a,} Recorded minimum eigenvalues during a full measurement run at a ``generated''mean photon number of 0.569, together with GBS data at 2.152 for comparison. 
    Each data point corresponds to a 1\,s integration window; the grey band denotes the uncertainty. 
    \textbf{b,} Highest and lowest observed minimum eigenvalues for generated GBS, SBS, and TBS data across all measured mean photon numbers. 
    Shaded regions represent uncertainties of reported values and have been connected to guide the eye. Uncertainties are estimated from counting errors.}
    \label{Fig:PaQSResults}
\end{figure}

We apply this measure to our experimental datasets, as summarized in Figure~\ref{Fig:PaQSResults}. 
Panel~\textbf{a} shows the minimum eigenvalues obtained for TBS, GBS, and SBS data generated in a single measurement run with a mean photon number of $0.569\pm0.011$, together with GBS data taken at a higher mean photon number of $2.152\pm0.090$. 
During each $\sim$11‑minute run, the system alternates between GBS and SBS/TBS configurations every 20\,s to mitigate long‑term drifts. 
Minimum eigenvalues are computed from 1\,s data blocks, each encompassing one million samples. 
This temporal segmentation enables investigation of possible time‑dependent effects, which may arise due to drifts in the phases of the input states. As expected, we find that only the GBS data exhibits such a temporal dependence due to the phase‑space asymmetry of SMSV states - i.e. the recorded minimum eigenvalues vary between different measurement bins as the input phases evolve. Notably, the GBS data at $\langle n \rangle = 0.569$ shows that more than 80\% of the recorded minimum eigenvalues are negative — thereby verifying quantumness in the generated data, whereas at $\langle n \rangle = 2.152$, the system displays quantum correlations less than 2\% of the time.

To benchmark the effect of increasing the input energy between the various sampling regimes, Figure~\ref{Fig:PaQSResults}\,\textbf{b} plots the obtained minimum eigenvalues for GBS, SBS, and TBS as a function of mean photon number. The range of these values at each brightness is illustrated by plotting both the lowest and highest obtained minimum eigenvalues. We note that, at the highest generated mean photon numbers, PNR cutoff does lead to artificially lowered minimum eigenvalues for TBS and GBS, which means the slight revival in nonclassicality for GBS at higher mean photon numbers is likely a false feature. This effect is described in further detail in Appendix \ref{App:impactpnr}.
For GBS at $\langle n \rangle = 2.152$, these extrema correspond to the highest and lowest eigenvalues from the data in Figure~\ref{Fig:PaQSResults}\,\textbf{a}. 
Both GBS and SBS data exhibit clear quantum signatures, whereas TBS data, as expected, shows no violations of the classical bound.

\section{Discussion}


We begin by examining the dependence of the input state brightness on the measured degree of quantumness. For TBS (Figure \ref{Fig:PaQSResults} dark and light blue lines), the minimum eigenvalue increases with mean photon number and, at the highest brightness, rises above the nonclassicality bound by more than $13\,\sigma$. This behavior is expected: increasing brightness simply adds photons from a thermal distribution, increasing photon‑number variance. In contrast, the SBS data (Figure \ref{Fig:PaQSResults} dark and light green lines) contains minimum eigenvalues that continuously decrease with increasing input brightness. At the highest mean photon number recorded, SBS data violates the classicality bound by more than $36\,\sigma$, therefore, \textit{brighter input states are advantageous in the SBS configuration}. 

In the GBS configuration, however, increasing the brightness of the input SMSV states does not yield a stronger or more statistically significant violation of the classical bound beyond a certain limit (Figure \ref{Fig:PaQSResults} dark orange line). The maximum violation exceeds $26\,\sigma$ at an average photon number of $\langle n \rangle = 0.172$, while the local minimum at $\langle n \rangle = 0.569$ corresponds to a $>14\,\sigma$ violation. Once the mean photon number exceeds the value at this local minimum, the system’s quantumness no longer continues to increase. Furthermore, Figure \ref{Fig:PaQSResults} reveals that at a mean photon number of $\langle n \rangle = 2.152$, the two-mode correlators from the GBS data do not exhibit quantumness for over 90\% of the input phase values. Therefore, \textit{using classical resources alone}, one could reproduce this data.

Understanding the cause for the observed reduction in quantumness is paramount. One natural hypothesis is that the reduced quantumness results from increased input-state impurity caused by system losses, consistent with recent work that decomposes impure SMSV states into a classical thermal component and a pure squeezed component \cite{Oh2024Alg}. Simulations performed using \textit{The Walrus}~\cite{Gupt2019} library, however, reveal that this behavior is compatible with simulations based on a lossless system (see Appendix \ref{App:simulations}), asking for explanations beyond input state impurity. The simulations were run with fixed input phases for each run and were calculated up to a photon number of 15---also ruling out the possibility that the reduction in quantumness stems from phase drifts or limited PNR. In addition to this, recent investigations into the impact of phase noise on the complexity of GBS have shown that the complexity is largely unaffected by phase noise \cite{paryzkova2025}.

We therefore postulate an alternative source for the observed behavior---that as the average photon number of the input states increases, the entanglement present at the output of the system begins to reside in higher-order correlations. Similar behavior has been observed in investigations into nonclassicality using comparable moment-based approaches \cite{Sperling2013,krishnaswamy2026}. However, this cannot be investigated in the current experimental setup as the reliable analysis of higher-order correlators would require experimental PNR resolution of greater than 4. Initial theory investigations in this direction indicate that the correlations can be found in higher-order metrics, but require further investigation and are therefore deferred to future work. Interestingly, similar behavior is not observed in the SBS data, which we postulate is due to the photon-number heralding collapsing the photon-number superposition present in GBS, thereby defining a well known input photon-number distribution. Although the precise mechanism remains unresolved, our findings demonstrate firstly that \textit{SBS provides advantages in our experimental implementation} -- a PNR cutoff of 3 is sufficient to verify nonclassicality using our metric and, in contrast to GBS, the nonclassicality using this metric increases with brighter input states. These results indicate that GBS with brighter input states is, in comparison, more difficult to implement -- consistent with conclusions from previously reported results that the total variation distance of lossy GBS systems worsens with brighter inputs \cite{Yu2023}. These results demonstrate that modifying the input state can yield markedly different system behavior and introduces a new framework for benchmarking emerging platforms aimed at extending the computational capabilities of sampling systems.

\begin{acknowledgments}
M.S would like to acknowledge many fruitful discussions with various colleagues and friends: Sevag Gharibian, Dhruva Sambrani, Hamid Naeij, Tina Chen, Dorian Rudolph, Jonas Kamminga, Robert Schade, Nathan Walk, Helen Chrzanowski, Jens Eisert, Dario Cilluffo, Martin Plenio, Andrew White, Fatemeh Mohit, Craig Hamilton, Shivani Singh, Aur{\`e}l G{\`a}bris, Magdalena Paryzkova, Igor Jex, Tim Ralph, Ryan Marshman, Nibedita Swain.

F.L would like to thank Laura Serino for help with the SLM.

F.S. is part of the Max Planck School of Photonics supported by the Dieter Schwarz Foundation, the German Federal Ministry of Research, Technology and Space (BMFTR), and the Max Planck Society.

T.S and T.B acknowledge partial funding by the European Union (ERC, QuESADILLA, 101042399). Views and opinions expressed are however those of the author(s) only and do not necessarily reflect those of the European Union or the European Research Council Executive Agency. Neither the European Union nor the granting authority can be held responsible for them.

This work has received funding from the German Federal Ministry of Research, Technology and Space within the PhoQuant project (Grant No. 13N16103).
\end{acknowledgments}

\subsection*{Author contributions}
The research project was initiated and directed by C.S., who oversaw its conceptual development and execution.
M.S. and K.-H.L. designed the experiment with input from B.B, and C.S.;
K.-H.L. and S.A. built the experimental setup with the help of F.L., J-L.E., C.P., F.Sc., J.L., M.S., M.Y., A.K., C.S., M.L., O.K., and F.St.;
M.Y., A.K., and R.H. developed and/or partially installed the laser system and path length compensation units;
F.Sc. designed and constructed the multiplexed detection scheme for heralding with aid from J.L.;
M.Y., S.A., K.-H.L. implemented the PLCU scheme;
F.Sc., T.S., T.B., J.L. developed and implemented the PNR detection scheme;
F.L. implemented the pulse shaper with help from K.-H.L., and F.Sc.;
K.-H.L. designed and built the squeezed-light source with the help of F.L., J-L.E., and F.Sc.;
J.-L.E., K.-H.L., S.A, M.R. performed the experimental runs and data collection with the help of J.L. and C.P.;
J.-L.E. completed data analysis with the help of L.A., J.S., J.L., and C.P.;
M.S. prepared the manuscript with the help of K.-H.L., S.A., J-L.E., F.L., C.P., F.Sc., L.A., J.S., M.R., B.B., C.S., and J.L.;
All authors discussed the results and reviewed the manuscript.

\subsection*{Data Availability}
All data are available in the manuscript, or online at https://doi.org/10.5281/zenodo.19469168.

\subsection*{Competing interests}
The authors declare no competing interests.

\subsection*{Correspondence and requests for materials} 
Request should be addressed to Michael Stefszky (stefszky@mail.uni-paderborn.de), Jan-Lucas Eickmann (janlucas@mail.uni-paderborn.de), or Christine Silberhorn (christine.silberhorn@uni-paderborn.de).

\appendix

\section{Review of Sampling Architectures}

\label{App:Review}

We provide a brief review of the SBS, TBS and GBS schemes due to their importance in the presented work. In particular, we highlight the relationship between these two sampling schemes and how we expect them to differ in the presented measurements. 

Boson sampling schemes were first introduced as a means of demonstrating quantum advantage in the current noisy intermediate-scale quantum era. 
In the original scheme \cite{Aaronson2010}, now referred to as simply boson sampling or Fock-state boson sampling, one inserts single-photon Fock-states $|1\rangle$ into a number of modes $K^2 < M$ of the interferometer, performs an arbitrary unitary rotation of the mode basis $\mathbf{\hat{U}}$, and finally measures the probability of photon-number outputs $P(m_1, \dots, m_M)$. 
Due to the indistinguishability of bosons, all possible permutations of a photon starting in mode $i$ and arriving in mode $j$ interfere. 
This corresponds to calculating the permanent of the matrix $\mathbf{U}_S$ 

\begin{equation}
    P(m_1, \dots, m_M) = \frac{|\mathrm{Per}(\mathbf{U}_S)|^2}{m_1! \dots m_M!}
\end{equation}
where $\mathbf{U}_S$ is constructed from $\mathbf{\hat{U}}$ depending on where and how many photons enter and exit the interferometer \cite{Kruse2019}. 
The permanent is strongly believed to be $\#P$-hard, making the boson sampling platform a suitable candidate for demonstrating quantum advantage \cite{Valiant1979}.

The first experimental implementations of boson sampling were limited to small numbers of photons due to the challenge of building reliable and scalable Fock-state sources \cite{Broome2013, Spring2013, Tillmann2013, Crespi2013}. In following implementations, the probabilistic single-photon sources have been replaced with deterministic single-photon sources, however, the scalability of these sources is still limited \cite{Loredo2017,Wang2019}. 
To improve the scalability of the system, what is now referred to as scattershot boson sampling (SBS) was proposed \cite{Lund2014} and experimentally implemented \cite{Bentivegna2015, Zhong2018}. In these early implementations heralding was limited to single-photon detection, although in principle one could use PNR detectors to enable SBS with higher-order Fock states.
Conceptually, one uses the photon-number correlations characteristic of a two-mode squeezed vacuum states (TMSV) in order to herald the number of inserted photons at each individual mode of the interferometer, thereby lifting the requirement to produce $N$ single photons in the first $K$ modes of the interferometer at the cost of allowing different input photon-number distributions.
Specifically, one utilizes multiple TMSV states of the form
\begin{equation} \label{eq:tmsv}
    \ket{\Psi_i} = \sqrt{1 - |\lambda_i|^2} \sum_{n = 0}^{\infty} \lambda_i^n \ket{n}_{\mathrm{herald}, i} \otimes \ket{n}_{\mathrm{signal}, i}
\end{equation}
involving interferometer (signal) and heralding (herald) modes $i$ respectively, where $\lambda_i \in \mathbb{C}_{< 1}$ is related to the squeezing parameter $r_i$ via $\lambda_i = \mathrm{tanh}(r_i)$. 
By post-selecting on heralding patterns $(h_1, \dots, h_K)$ where $h_i$ photons are detected in the heralding mode $i$, one can build up the conditioned probabilities $P(m_1, \dots m_M | h_1, \dots, h_K)$ of detecting $m_j$ photons in the interferometer output mode $j$ given the heralding pattern
\begin{equation}
    P(m_1, \dots m_M | h_1, \dots, h_K) = \frac{|\mathrm{Per}(\mathbf{U}_S)|^2}{m_1! \dots m_M! h_1! \dots h_K!}
\end{equation}

If one collects statistics for one particular heralding pattern $(h_1, \dots, h_K) = (1, \dots, 1)$, the SBS protocol becomes equivalent to the original boson sampling protocol with an overhead in runtime \cite{Keshari2015}. 
Although scattershot boson sampling eliminates one requirement, scaling the system remains a problem as the probability for detecting a specific heralding pattern
\begin{equation}\label{eq:sbs_prob}
    P(h_1, \dots, h_K) \propto \prod_{i =1}^{K}|\lambda_i|^{2 h_i}
\end{equation}
decreases exponentially with the number of heralded photons. We note that our PNR heralding scheme enables the first demonstration of SBS with higher-order Fock states.

Due to the limited scalability of Fock-state boson sampling and SBS, new methods for scaling were investigated. In particular, people began to explore the possibility of replacing the Fock state inputs with deterministically generated input states. In early investigations, the complexity of thermal boson sampling (TBS) -- in which the input states are set to be thermal states, was explored \cite{Keshari2015}. This multimode input state can be prepared by tracing over one mode from each input channel of the multimode TMSV shown in Equation \ref{eq:tmsv},
\begin{equation} \label{eq:therm}
    \hat{\rho}_i = \sqrt{1 - |\lambda_i|^2} \sum_{n = 0}^{\infty} \lambda_i^n \ket{n}\bra{n}
\end{equation}

It was shown that driving the system with thermal states of equal mean photon number (or equivalently temperature $\mu$) trivially leads to an output state with no correlations. Furthermore, it was shown that calculating the probability of a particular photon-number output in the case of thermal states with different mean photon number requires calculating the permanent of a submatrix $\mathbf{U}_T$ of the complete unitary operation. In contrast to SBS, this submatrix is determined only by the locations where photons exit the interferometer.
\begin{equation}
    P(m_1, \dots m_M) = \left( \prod_{i=1}^{K} \mu_i \right)\mathrm{Per}(\mathbf{U}_T)
\end{equation}
Critically, the use of thermal input states leads to the matrix $\mathbf{U}_T$ being a positive-semidefinite Hermitian matrix. It was subsequently shown that Stockmeyer's approximate counting algorithm can be used to efficiently approximate the permanent of such matrices \cite{Keshari2015}. Therefore, sampling from the output probability distribution of TBS is not \#P-hard and an efficient classical algorithm exists that produces samples from this output probability distribution.

This led researchers to consider the use of single-mode squeezed vacuum (SMSV) states as input states \cite{Keshari2015, Huh2015}, a sampling scheme now referred to as Gaussian boson sampling (GBS). The complexity of this scheme was proven shortly thereafter \cite{Hamilton2017}, with subsequent work further elucidating regimes where complexity can be maintained \cite{Ehrenberg25,Shou25}. SMSV states can be generated deterministically via parametric down-conversion (PDC) processes, enabling rapid scaling of the size of experimentally implemented systems.

\begin{figure}[!ht]
    \centering
    \includegraphics[width = 0.7\textwidth]{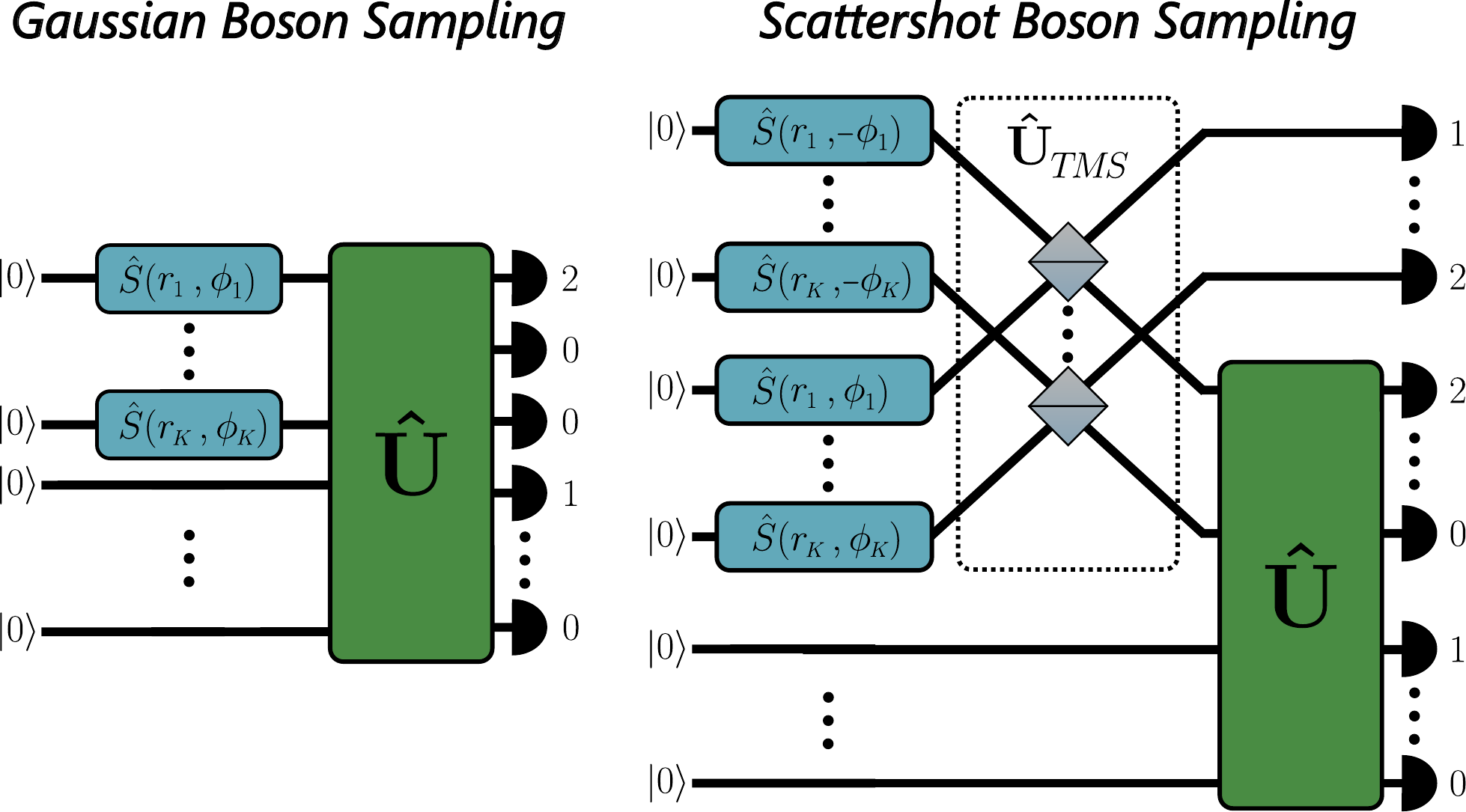}
    \caption{Equivalence of Gaussian and scattershot boson sampling. In Gaussian boson sampling (left) $K$ single-mode squeezed states are prepared and sent into an $M \geq K$  mode interferometer before measuring the photon-number at each output mode.
    Scattershot boson sampling (right) can be considered a particular instance of Gaussian boson sampling, where $K$ two-mode squeezed states are first prepared by interfering $2K$ single-mode squeezed states, shown by the transformation $\mathbf{\hat{U}}_{TMS}$. One mode of each two-mode squeezed state is used to herald on a photon-number input, while the other mode is sent into the interferometer to perform boson sampling.}
    \label{Fig:GBS_SBS}
\end{figure}

More precisely, it was shown that generating GBS samples in a classical computer requires one to calculate matrix Hafnians, which, in general, are also strongly believed to be $\#P$-hard \cite{Hamilton2017}. 
Gaussian states are fully characterized using their $2M\times2M$ covariance matrix $\sigma$ and a displacement vector $d$, which in the original GBS scheme was assumed to be zero $d=0$. 
In this case, the probability of measuring $m_j$ photons in mode $j$ after the interferometer for a non-displaced Gaussian state ($d = 0$) is
\begin{equation}
    P(m_1, \dots, m_M) = \frac{\mathrm{Haf}(\mathbf{A}_s)}{m_1! \dots m_M!\sqrt{\mathrm{det}\left(\sigma + \mathbb{I}_{2M}/2\right)}}
\end{equation}
where $\mathbf{A}_s$ is a sub-matrix of the matrix $\mathbf{A} = \begin{pmatrix}0 & \mathbb{I}_M\\ \mathbb{I}_M & 0\end{pmatrix} (\mathbb{I}_{2M} - \sigma^{-1})$. 
In contrast to Fock-state boson sampling, the sub-matrix $\mathbf{A}_s$ is constructed solely from the detected output pattern, which originates from no longer heralding on a specific input pattern, or equivalently, that the input is a coherent superposition of all $N$-photon patterns from the Gaussian input. 
It can be shown that using classical states at the input results in specific cases exhibiting reduced complexity. In the case of thermal states the Hafnian (or using a matrix identity - the permanent) can be simplified and a classical algorithm can be found \cite{Tamma2014,Keshari2015,Kruse2019}, whereas for coherent inputs ($d \neq 0$ and $\sigma = \mathbb{I}_{2M}$) the problem reduces to simply sampling from $M$ separate Poissonian distributions\cite{Kruse2019, Keshari2015}.
An important thing to note is that one can convert between SMSV states  and TMSV states by interference, thereby allowing one to view SBS as a subset of GBS as shown in Figure \ref{Fig:GBS_SBS}. 
TMSV states can be generated using multiple SMSV states via an initial interference layer, which can then be used to perform SBS.  
This relation between SBS and GBS can also be seen mathematically, since the Hafnian directly relates to the permanent of a matrix $\mathbf{C}$ \cite{Hamilton2017, Kruse2019}
\begin{equation}
    \mathrm{Per}(\mathbf{C}) = \mathrm{Haf}\begin{pmatrix}\mathbf{0} & \mathbf{C}\\\mathbf{C}^T & \mathbf{0}\end{pmatrix}.
\end{equation}
Despite this connection between SBS and GBS, there are major differences worth highlighting. 
While the input state generation for SBS is probabilistic, as shown in Equation\ref{eq:sbs_prob}, the squeezed state generation in GBS is deterministic. 
Equivalently, \textbf{GBS utilizes the full PDC state, a coherent superposition of photon-number states}, which is the source of the improved scaling offered by this system. 
Furthermore, the matrix $\mathbf{A}$ that is sampled in GBS depends on the squeezing strength $r_j$ and the phase angle $\phi_j$ of the contributing SMSV states. 
In the case of SBS neither the phase nor the strength of the TMSV states contribute beyond the heralding probability factor into the probability distribution. 
Therefore, GBS offers a much richer space of interference dynamics, and provides additional degrees of freedom, which serve as encoding parameters when investigating possible applications.

\section{Experimental Design}

\label{App:expdesign}

\subsection{Laser Preparation}

\subsubsection{Pulse Picking}
Our temporally-multiplexed, waveguided squeezed-light source requires a high degree of programmability and  optimization in order to produce the required quantity and quality of squeezed-light pulses.
Generation of temporal bins from the pump requires exact control over the timing and number of pulses arriving at the source for each measurement run, while precise tailoring of the spectral characteristics within each pulse ensures generation of indistinguishable, single spectro-temporal-mode TMSV states.

The entire system is driven by a (MenloSystems SmartComb) frequency comb source that produces femtosecond pulses in the telecom-band with a repetition rate $f_{\mathrm{RR}}$ = 80\,MHz. 
Both the carrier-envelope offset frequency \(f_{\mathrm{CEO}}\) and the repetition rate are actively stabilized to an RF oscillator. 
Following this stage, a high-power amplifier boosts the average output power to several watts before the beam enters a final SHG stage. 
The generated second harmonic is centered at around 772.5\,nm with a full-width at half maximum bandwidth of approximately 1\,nm and has a maximum average power of approximately 900\,mW.

Next, the pump pulse train is pulse-picked using a free-space pulse picker (QUBIG GmbH HVOS-NIR-1.5k100). The pulse picker is driven at a frequency of 1\,MHz, effectively reducing the repetition rate by a factor of 80. 
This driving frequency defines the high sampling rate of our system. 
By varying the time window in which the pulse picker is configured to pick pulses we are able to select a desired number of consecutive pump pulses for pumping the nonlinear waveguide. 
Combined with the temporal demultiplexing of our squeezing source (see section \ref{subsec:demux}), this allows control over the number of squeezed states at the input of our interferometer.

\subsubsection{Spectral Shaping}

The next step in preparing the pump field is spectral shaping, which ensures that the generated squeezed states are spectro-temporally decorrelated, thereby ensuring optimal interference between different pulses. 
This shaping is achieved using a standard pulse shaping setup: The incoming pulses are sent onto a grating with groove spacing ${1}/{2200}$\,mm, resulting in angular separation of the spectral components.
Thereafter, a cylindrical lens, with focal distance $200$\,mm, converts the angular separation of the spectral components to a spatial separation, while also horizontally focusing each spectral component onto the spatial light modulator (SLM, Santec SLM200). 

\begin{figure}[!ht]
    \centering
    \includegraphics[width=0.8\textwidth]{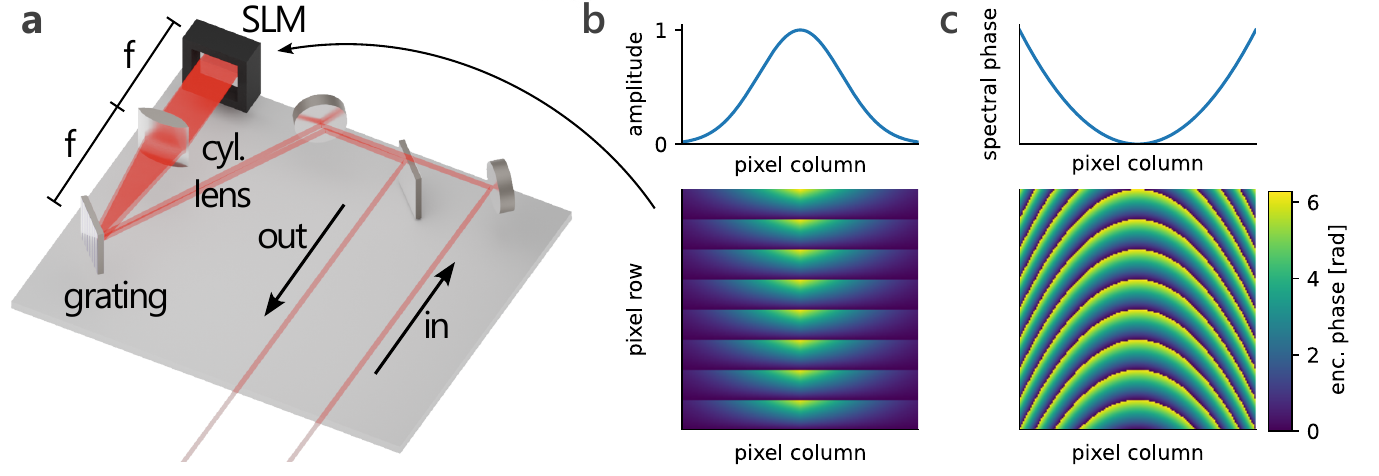}
    \caption{Overview of pulse shaper design and implementation.
        \textbf{a} Experimental setup of the pulse shaper. It consists of a zero-dispersion line, containing a grating for spectral separation and a cylindrical lens to focus spectral components onto the spatial light modulator (SLM). 
        \textbf{b} Example encoding of Gaussian spectral amplitude on the SLM. Shown are the desired amplitude (top) and encoded phase-mask on the SLM (bottom). 
        \textbf{c} Example encoding of a quadratic spectral phase.
        Shown are the desired spectral phase profile (top) and encoded phase mask (bottom).
    }
    \label{fig:pulse_shaping_setup}
\end{figure}

Each vertical slice at position $x$ of the SLM can be used to program a specific frequency component $\omega(x)$ of the incoming field. 
For each, the SLM encodes a tunable blazed grating, providing full programmability of the amplitude and phase of individual frequency components (see Figure \ref{fig:pulse_shaping_setup} \textbf{b} for example masks).
As the field propagates backwards though the setup the spectral components are then recombined into one spatial mode.
Using this setup, the spectral width and phase required for optimal state generation is found experimentally by maximizing the second-order auto-correlation value $g^{(2)}$ measured in one mode of a generated TMSV state (more details on measuring $g^{(2)}$ in section \ref{subsec:pdc_source}).

\subsection{Squeezed-Light Source}
\label{subsec:pdc_source}

The squeezed states in the PaQS system are generated in a scheme previously unexplored in GBS systems -- multiple squeezers are generated temporally in a waveguided, single-pass type-\MakeUppercase{\romannumeral 2} nonlinear material. 
Generating multiple squeezers temporally overcomes the problem of producing and controlling multiple identical waveguide samples and provides a more resource efficient method to generate multiple squeezers. Furthermore, this scheme generates ps long pulses of squeezed-light that are fully compatible with integrated optics and the implemented PNR scheme.
To ensure that subsequently generated squeezers completely interfere with one another requires a high degree of spectro-temporal decorrelation, while the ability to transform the generated state between SMSV and TMSV requires a high degree of spectro-temporal indistinguishability between the generated signal and idler fields.

The integrated quantum light source is a 20\,mm-long periodically poled potassium titanyl phosphate (PPKTP) waveguide (AdvR Inc.) with $3.5\,\mu$m waveguide width, engineered to maintain single-spatial-mode operation at 1550\,nm for both polarizations. The dispersion properties of KTP allow it to generate decorrelated and spectro-temporally indistinguishable photons at telecom wavelength in a type-\MakeUppercase{\romannumeral 2} PDC process \cite{Eckstein2011,Harder2013}. 
The waveguide exhibits a propagation loss of $0.25 \pm 0.03$\,dB/cm. The total loss experienced by the state exiting the source is approximately $13 \pm 1$\,\%, which arises from propagation losses through half of the sample and the Fresnel reflection on the uncoated end-facet of the waveguide.

Neglecting losses, the state generated in the PDC process is given by
\begin{equation} \label{eq:PDC_state}
    \ket{\Psi_{\mathrm{TMSV}}} = \sqrt{1 - |\lambda|^2} \sum_{n = 0}^{\infty} \lambda^n \ket{n}_{\mathrm{H}} \otimes \ket{n}_{\mathrm{V}}
\end{equation}
corresponding to one of the many TMSV states required for SBS or TBS (cf. Equation \eqref{eq:tmsv}). 
The labels $\mathrm{H}$ and $\mathrm{V}$ refer to horizontal and vertical polarization, respectively, representing the two-modes of the TMSV state.
One can deterministically split the two polarizations into different spatial modes using a polarizing beam splitter (PBS), thereby allowing one to implement SBS or TBS configurations. 

For implementing GBS, one has to prepare input states of the form
\begin{equation} \label{eq:SMSV}
    \ket{\Psi_{\mathrm{SMSV}}} = \frac{1}{\sqrt{\mathrm{cosh}(r)}} \sum_{n = 0}^{\infty} \biggl(-\frac{e^{i\phi} \mathrm{tanh}(r)}{2}\biggr)^n \frac{\sqrt{(2n)!}}{n!} \ket{2n},
\end{equation}
which can be achieved by interfering the two modes of the TMSV states. 
By rotating the polarization of both signal and idler fields by 45 degrees, the PBS acts as a 50:50 BS and the state exiting the two ports is described by two independent SMSV states
\begin{equation}
    \ket{\Psi_{\mathrm{out}}} = \ket{\Psi_{\mathrm{SMSV}}}_H \otimes \ket{\Psi_{\mathrm{SMSV}}}_V.
\end{equation}
To enable fast switching between the two states, this polarization rotation is implemented using an electro-optic modulator (EOM). 
In the off state (where 0\,V applied to the internal high-voltage amplifier), the EOM implements no polarization rotation and the photons are split up deterministically. When driving the EOM amplifier with a voltage of $0.68$\,V (corresponding to $V_\pi/2$ and approximately 700\,V applied to the EOM) the signal and idler fields are interfered on the PBS. 
Note that for both TMSV and SMSV, the mean photon-number per mode of the state relates to the squeezing parameter $\langle n \rangle = \mathrm{sinh}^2(r)$ and is the same in both cases.

To confirm our source meets the stringent requirements of spectro-temporal decorrelation and indistinguishability, we perform two key measurements: second-order auto-correlation of the generated state is used to confirm decorrelation \cite{Christ2011} and HOM interference between the two polarization components confirms indistinguishability between the signal and idler fields.

\subsubsection{Second-order auto-correlation}
\label{subsec:second-order}

The second-order auto-correlation $g^{(2)}$ is extracted from the experimental scattershot boson sampling run presented in the main paper. 
For all measurements, a 2\,nm spectral filter is placed at the output of the waveguide, the spectral bandwidth of which is chosen to remove the sinc side-lobes that typically arise in PDC while minimally affecting the central peak of the phasematching profile. The photon-number statistics for all eight pulses (input modes) in the heralding arm are collected throughout the whole measurement run. From the photon-number statistics the $g^{(2)} = \frac{\sum_np_nn(n-1)}{(\sum_np_nn)^2}$ for each separate mode/input is calculated and the average over all inputs is taken.
This treatment is repeated for all of the implemented squeezing parameters.

It can be seen that, as the system approaches lower powers, the $g^{(2)}$ increases rapidly. 
This is due to imperfect splitting of the signal and idler fields on the PBS that leads to some small amount of interference due to the spectro-temporal indistinguishability of signal and idler fields, and therefore some single-mode squeezing in the mode. 
The second order correlation function of a SMSV state is given by $g^{(2)}_\mathrm{SMSV}=3+\frac{1}{\langle n \rangle}$ which diverges to $\infty$ as $\langle n \rangle\to0$. 
Due to this effect, the second-order correlation function value used to determine the degree of correlation in the source is taken at a mean photon-number of $\langle n \rangle \approx 0.57$. 
At this mean photon-number the contribution of the SMSV is negligible, resulting in  $g^{(2)} = 1.95\pm0.03$, from which the effective number of spectro-temporal modes $K$ is  determined to be $K = \frac{1}{g^{(2)}-1} = 1.05\pm0.03$ \cite{Christ2011}.

\subsubsection{Signal and Idler Indistinguishability}

A high degree of spectro-temporal indistinguishability between the generated signal and idler fields of the PDC state is required in order to effectively transform the TMSV states exiting the waveguide into SMSV states. To quantify the spectro-temporal indistinguishability between signal and idler, the voltage of EOM2 is varied between SMSV and TMSV operation modes, recording coincidences $n_C$ for each step.
This variation interpolates between the full polarization distinguishability of the TMSV case and full indistinguishability of the SMSV case.
As such, one can consider this second measurement as a HOM interference experiment, where distinguishability is varied through polarization manipulation rather than delay scanning.

To map the visibility from the polarization scanning method to the standard time-delay HOM visibility \cite{HOM1987, HOM2024} one can calculate it via $V_{HOM, pol} = \frac{0.5\cdot n_{C, max} - n_{C, min}}{0.5\cdot n_{C, max}}$ \cite{Meyer-Scott2018}. 
The reported visibility of $V_{HOM, pol} = 96.3\% \pm 1.2\%$ is extracted by performing a sinusoidal fit to the measured coincidence counts, demonstrating the high degree of indistinguishability between the signal and idler photons. A difference in maximum coincidence counts between the two possible TMSV configurations was observed: HV $\to$ HV and HV $\to$ VH, at approximately 0\,V and 1.3\,V respectively.
This effect stems primarily from imperfect spatial overlap between the waveguide's orthogonal polarization modes, which are coupled to single-mode fibers.

\subsection{Demultiplexer} \label{subsec:demux}
\begin{figure}[!ht]
    \centering
    \includegraphics[width = 0.8\textwidth]{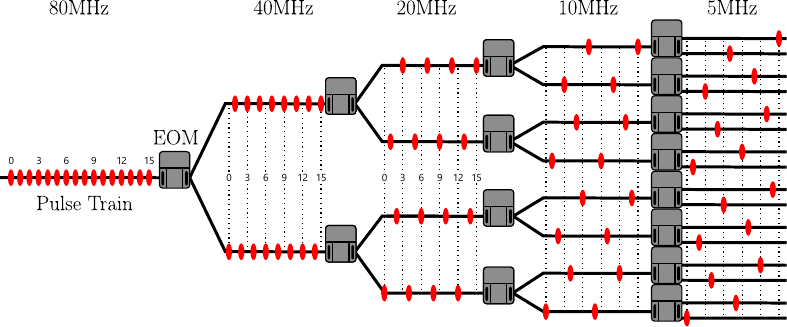}
    \caption{Concept of the EOM splitting tree for translating from time into spatial bins. A pulse train with a repetition rate of $80$\,MHz and $\leq 16$ pulses enters the EOM tree. At each EOM the repetition rate is halved, such that after four steps at each of the outputs a repetition rate of $5$\,MHz would be achieved, which for the given input configuration corresponds to at maximum a single pulse at each output.}
    \label{Fig:Demux_Concept}
\end{figure}

With the desired number of $k$ optimized squeezed-light pulses exiting the waveguide in a $k$-mode pulse train, the next step is to demultiplex these pulses from the time domain into the spatial domain so that they can be inserted into the spatially implemented interferometer. 
To achieve this we use a demultiplexer (Qubig DMX16\textunderscore80) that routes up to 16 individual pulses into separate polarization-maintaining single-mode fibers with a specified average transmission of $\geq 80$\,\%. 
The routing is implemented in a 4-level tree of electro-optic modulators (EOMs) that effectively halve the 80\,MHz repetition rate at each step, as shown schematically in figure \ref{Fig:Demux_Concept}.

\begin{figure}[!ht]
    \centering
    \includegraphics[width = 0.8\textwidth]{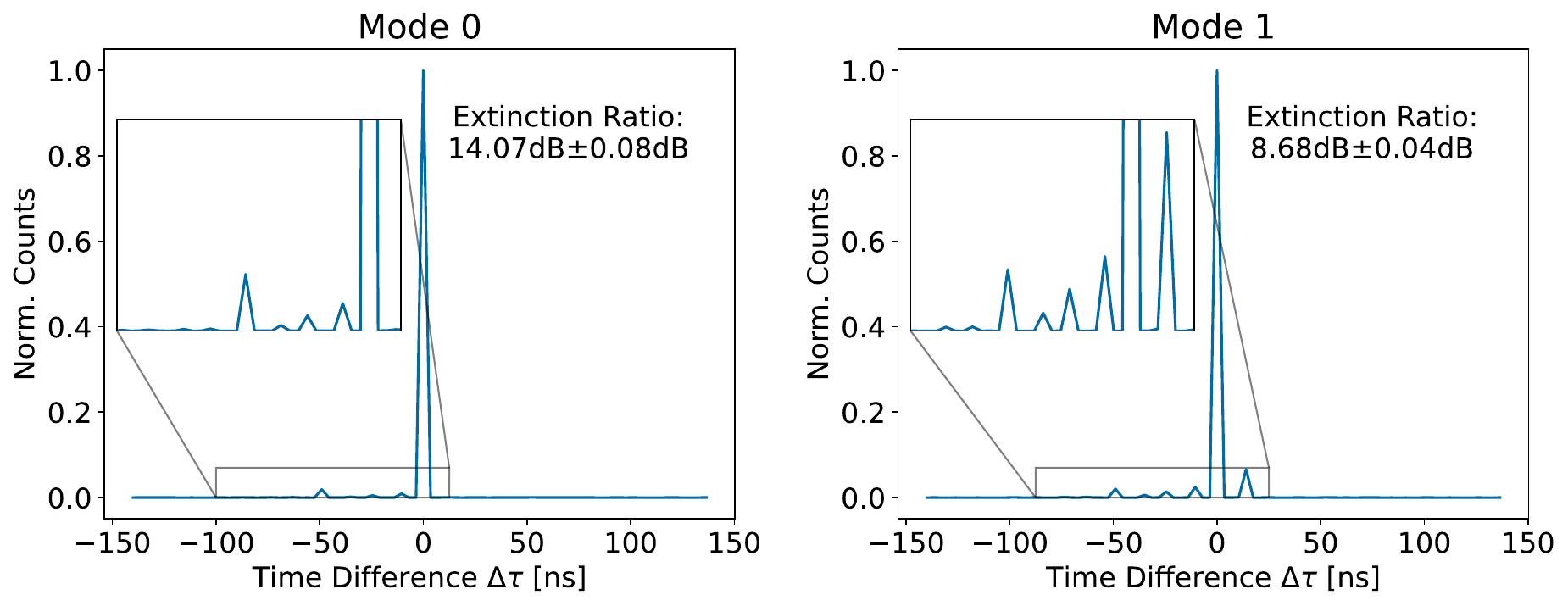}
    \caption{Example for the extinction ratio measurements of the first two input modes. Wrongly routed photons arrive with a time delay relative to the correctly routed photons.}
    \label{Fig:Demux_ER}
\end{figure}

The extinction ratio of the device characterizes the amount of light that is routed from undesired input modes to any given output mode. 
When photons exit in the incorrect fiber, they arrive with a time delay of $\Delta\tau = k\cdot 12.5$\,ns, $k \in \mathbb{Z}^*$ relative to the correctly routed pulse. 
By measuring the photon counts relative to the laser trigger we can thereby identify the eight different pulses and calculate the extinction ratio $ER$ for each mode 
\begin{equation} \label{eq:ER}
ER = 10\cdot \mathrm{log}_{10}\biggl(\frac{\mathrm{Counts}(\Delta\tau = 0)}{\sum_{\Delta\tau\neq0}\mathrm{Counts}(\Delta\tau)}\biggr).
\end{equation}
An example for such a measurement can be seen in figure \ref{Fig:Demux_ER}, the resulting extinction ratios for all modes are collected in Table \ref{tab:ER}.

Since data acquisition happens in a narrow window of $\ll$12.5\,ns, the incorrectly routed photons will be filtered out at the detection stage and therefore act only as a source of system loss, and not crosstalk, which would further reduce the measured Klyshko efficiencies by introducing non-correlated photons \cite{Krapick2013}. 

\begin{table}[!ht]
  \centering
    \begin{tabular}{|l||c|c|c|c|c|c|c|c|}
    \hline
    \textbf{Mode} & $0$ & $1$ & $2$ & $3$ & $4$ & $5$ & $6$ & $7$ \\
    \hline
    $\mathbf{ER}$ \textbf{[dB]} & $14.07$ & $8.68$ & $10.24$ & $10.57$ & $12.16$ & $9.31$ & $11.39$ & $10.97$ \\
    \hline
    $\Delta\mathbf{ER}$ \textbf{[dB]} & $0.08$ & $0.04$ & $0.06$ & $0.05$ & $0.06$ & $0.05$ & $0.05$ & $0.05$ \\
    \hline
    \end{tabular}
    \caption{Demultiplexer Extinction Ratios. Measured extinction ratios $\mathrm{ER}$ (Equation \ref{eq:ER}) and uncertainties $\Delta \mathrm{ER}$ for the $8$ used output channels of the demultiplexer. The extinction ratios are measured using the technique described in \ref{subsec:demux}.}
  \label{tab:ER}
\end{table}

\subsection{Path Length Compensation System} \label{subsec:PLCS}
After the demultiplexer, each temporal bin $k$ now resides in a distinct spatial mode $k$ (i.e. PM fiber). 
However, the individual pulses are still delayed relative to one another in time by the initial separation $\Delta\tau = k \cdot 12.5$\,ns, $k \in \mathbb{Z}^*$. 
In order to guarantee interference between different pulses, it is necessary to ensure that all pulses are temporally overlapped at the interferometer. 
The temporal overlap is optimized in the path length compensation system (Menlo Systems GmbH). 

The system comprises $8$ path length compensation units (PLCUs) that are enclosed in a 10\,mm thick polyethylene box for thermal and acoustic isolation.
Each PLCU consists of a length of fiber wrapped around a piezo-actuated fiber stretcher and two Peltier elements enclosed in an aluminum box further insulated with polyethylene. 
The fiber stretcher includes 60\,m of standard polarization maintaining (PM) fiber (PM 1550, Corning Inc.) and has a specified modulation constant of approximately 33\,rad/V at a frequency up to 5\,kHz. 
The Peltier elements can be set between 15-35\,$^\circ$C with a precision of 0.05\,$^\circ$C. 
Full control of this system is provided by a Menlo Systems Syncro unit.

To compensate for temporal delay between pulses in different modes, a fiber length of roughly $L = k \cdot 2.56$\,m, corresponding to $\Delta\tau = k \cdot 12.5$\,ns, is spliced to the $k$-th unit. 
The unwanted optical path mismatch can be reduced to 0.5\,mm (i.e. 2.5\,ps) by several iterations of fiber splicing. 
The final optimization of the temporal overlap is achieved by tuning the temperature of the 60\,m of fiber in each PLCU. 
The thermal coefficient of delay for standard single-mode fiber is approximately 33-50\,ps/km/K, allowing each to be tuned by approximately 2\,ps/K. 
Given the range and precision of the temperature control inside each PLCU, it is possible to both perform a full scan of the HOM interference dip and set it to its optimal position, as demonstrated in Appendix \ref{App:expdesign}.

\subsubsection{Fiber Dispersion}
\label{subsubsec:fiberdisp}

One issue that arises in the temporally multiplexed squeezed-light source scheme is that, due to dispersion, the different lengths of fiber traversed by each pulse will lead to pulse broadening. 
This will lead to a reduction in the interference visibility between pulses, or equivalently, will introduce distinguishability between different input modes - with the effect becoming more pronounced as pulse duration is decreased.
Assuming two Gaussian pulses with initial pulse duration \(\tau_0\) (field RMS) traveling through a dispersive medium of lengths \(L_0\) and \( L_0 + \Delta L \), respectively, one can write the resulting electric fields,
\begin{align}
\tilde{E}_1(\omega) &= \tilde{E}_0(\omega) \exp\left[ -\frac{(\omega - \omega_0)^2 \tau_0^2}{2} (1 + iC) \right], \\
\tilde{E}_2(\omega) &= \tilde{E}_0(\omega) \exp\left[ -\frac{(\omega - \omega_0)^2 \tau_0^2}{2} \left(1 + iC + i \frac{ \Delta \phi'' }{ \tau_0^2 } \right) \right].
\end{align}
where $\tilde{E}_1(\omega)$ is the electric field amplitude, $\omega$ is the angular frequency, $\omega_0$ is the carrier (central) frequency, $\Delta \phi''= \mathrm{GVD} \cdot \Delta L$ is the quadratic phase accumulated due to group velocity dispersion (GVD) in \(\Delta L\), and $ C = \phi''/\tau_0^2$ is the common chirp with $\phi''= \mathrm{GVD} \cdot L_0$.
To determine the impact of the difference in quadratic phase $\Delta \phi''$ on the intensity visibility of the two interfering Gaussian pulses, we need to calculate the overlap integral:
\begin{equation} \label{eq:OverlapInt}
I = \int_{-\infty}^{\infty} \tilde{E}_1^*(\omega)\, \tilde{E}_2(\omega)\, d\omega.
\end{equation} 

After normalizing and solving \eqref{eq:OverlapInt}, and assuming that the pulses overlap perfectly in time, one arrives at the following formula for the intensity visibility,
\begin{equation} \label{eq:dispersion}
V = \frac{\tau_0^2}{\sqrt{\tau_0^4 + \left( \frac{\mathrm{GVD} \cdot \Delta L}{2} \right)^2}}.
\end{equation} 
For standard PM fibers at 1545 nm wavelength (GVD = -26\,fs$^2$/mm) with \(\Delta L_{\mathrm{max}} = 18 \)\,m (i.e. the approximate fiber length difference between PLCU units 0 and 7), and an initial pulse duration of \(\tau_0 \approx 1 \pm 0.1\)\,ps, the expected visibility is \( V \approx 0.974^{+0.008}_{-0.013}\). We therefore see that the impact of fiber dispersion in the current setup is minimal.

\subsubsection{Phase stability}
\label{subsubsec:phasestab}

\begin{figure}[!ht]
    \centering
    \includegraphics[width = \textwidth]{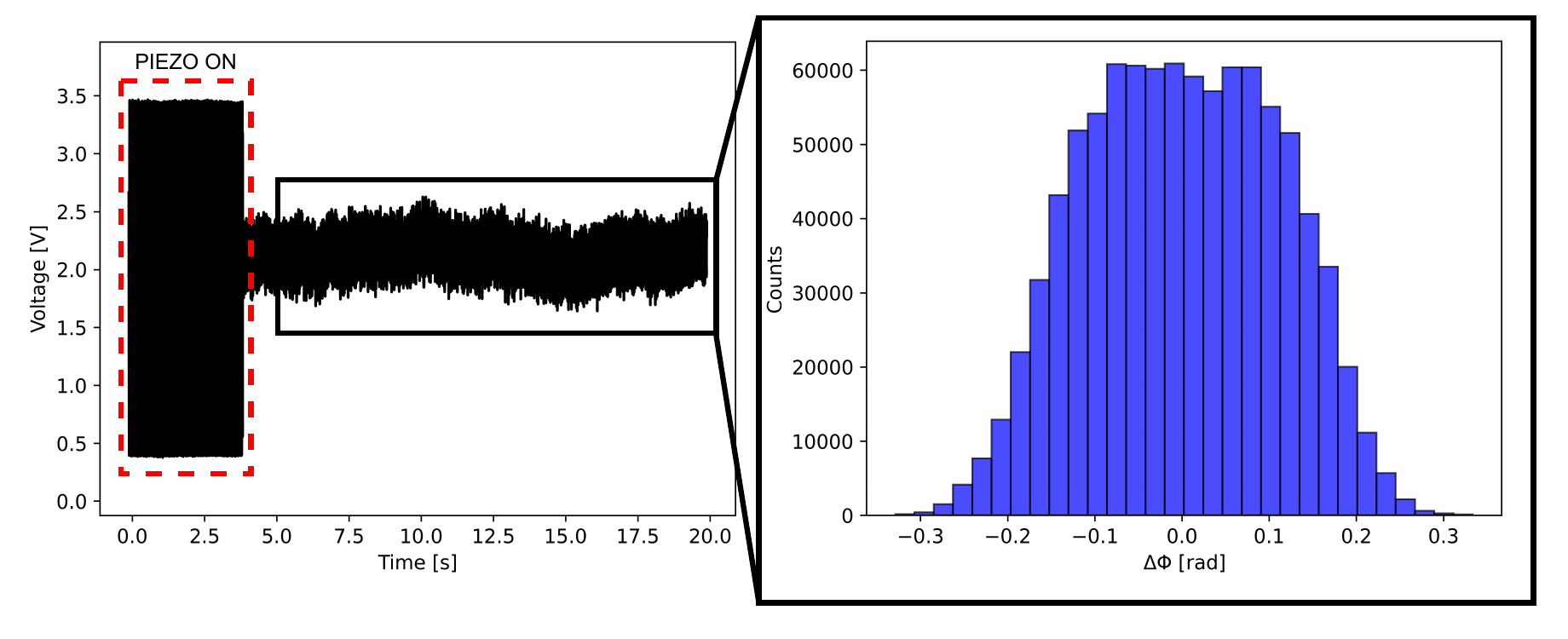}
    \caption{System phase stability measurement. As described in the text, we observe the voltage produced by a photodiode observing one output port of a BS interfering two coherent fields injected into two input modes of the interferometer. For approximately four seconds the piezo controller is actuated in order to provide a phase calibration. After this time, the actuation signal is removed and the system is left to drift. On the right, the phase distribution calculated from the last 15\,s of the data is presented. The rms value of the distribution is 0.11\,rad, while its FWHM is 0.33\,rad.}
    \label{Fig:PLSU_stability}
\end{figure}

The length of the fibers in the PLCUs has been minimized in order to reduce unwanted phase fluctuations, while still providing sufficient range for compensating the path length difference between different input pulses. 
The system has a high degree of passive isolation to thermal and acoustic influences due to multiple levels of shielding. 
The performance of this passive stabilization is measured by implementing a 50:50 BS within the interferometer between two input channels and injecting coherent light pulses from the telecom-band laser source into the system. 
The interference observed in one output of the implemented BS is then recorded by detecting the field intensity on a photodetector. 
The voltage detected in this way is then observed, first as the piezo in the PLCU is driven with a high frequency voltage, in order to provide a calibration reference, and then, after this voltage is switched off and the system is left to drift. 
The signal recorded in this way is shown in Figure \ref{Fig:PLSU_stability}. 
We see that the mean of the measured voltage does not change significantly over the 15 second measurement time after the driving voltage has been removed, indicating mean phase stability on this timescale. However, there is an appreciable variance on the measured signal, which can be related to a phase noise using the calibration reference. 
The inferred phase noise is 0.11\,rad, with a FWHM of 0.33\,rad. 
Multiple measurements taken in this way reveal that most measurements show a similar degree of stability, although drifts can sometimes be observed. 
The observed stability, combined with the fact that samples are generated at a rate of 1\,MHz, results in the ability to gather millions of samples with unknown input phases that nevertheless have a small variance and almost constant mean.

Although the passive stability of the system is very high, active stabilization is required in order to program the input phases and to run the system on timescales greater than multiple seconds with unchanging phases. 
The PLCUs do include the means to implement such a scheme, as they include both a piezo drive and Peltier elements for controlling the phase. 
This would require only derivation of a suitable error signal for the feedback, which is currently under development. 

\subsubsection{PLCU insertion losses}
\label{subsubsec:PLCUloss}

Finally, we report the measured insertion losses $IL$ for each unit in Table~\ref{tab:EL_PLCUs}. 
Unit 2 shows 1\,dB higher $IL$ than all the other PLCUs, most likely due to defective splicing in the connection to the fiber stretcher. 
The system remains fully functional despite this loss imbalance, which will be addressed in the next iteration of the experiment.

\begin{table}[!ht]
  \centering
    \begin{tabular}{|l||c|c|c|c|c|c|c|c|}
    \hline
    \textbf{Mode} & $0$ & $1$ & $2$ & $3$ & $4$ & $5$ & $6$ & $7$ \\
    \hline
    $\mathbf{IL}$ \textbf{[dB]} & $0.5$ & $0.8$ & $2.1$ & $0.5$ & $0.4$ & $1.0$ & $0.3$ & $0.9$ \\
    \hline
    \end{tabular}
    \caption{Measured insertion losses (IL) of the 8 PLCU units. Using a Thorlabs PM100D power meter and a S144C power meter head the insertion losses of the 8 PLCU units are determined using a telecom-band pulsed laser. The power meter head has a measurement uncertainty of 5\,\%.}\label{tab:EL_PLCUs}
\end{table}

\subsection{Integrated, Fully Programmable Interferometer }
\label{subsec:quix}

The spatially separated, synchronous light pulses are fed into a 12-mode fully programmable photonic processor (QuiX Quantum)\cite{Taballione} realized in the low-loss silicon nitride (Si\textsubscript{3}N\textsubscript{4}) platform using TriPleX technology\cite{TriPleX}.
The device features a two-dimensional network of temperature tunable Mach-Zehnder interferometers (MZIs), implementing the Clements configuration such that any $12\times12$ unitary transformation $U$ can be realized \cite{Clements}. 
The device allows one to program the desired unitary using a Python interface.

A standard measurement to quantify the performance of such interferometers can be performed by measuring how likely it is for input photons to scatter from a certain, known input mode $i$ to an output mode $j$ of the interferometer, known as the scattering probability $p_{i \rightarrow j}$.
After obtaining these probabilities for all combinations of modes $i$ and $j$ one can calculate the similarity (also referred to as fidelity, average state fidelity, transformation fidelity or amplitude fidelity) of the target implemented scattering $p_{i \rightarrow j}^{(t)} = |U_{ij}|^2$ to the experimentally measured scattering probabilities $p_{i \rightarrow j}^{(exp)}$ via 
\begin{equation} \label{eq:similarity}
\mathcal{S} = \frac{1}{N} \, \sum_{ij} \sqrt{p_{i \rightarrow j}^{(t)} p_{i \rightarrow j}^{(exp)}},
\end{equation}
where $N$ is the dimension/number of modes of the implemented unitary transformation.
It is important to note that this measure, as it is obtained from probabilities, is insensitive to the complex components of the implemented matrix.

\begin{figure}[!ht]
    \centering
    \includegraphics[width = 0.7\textwidth]{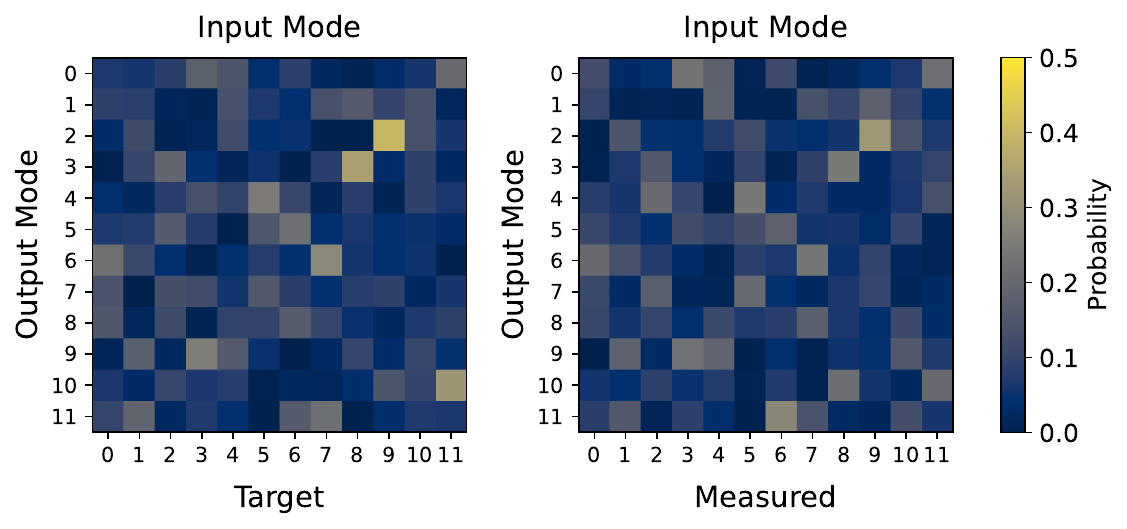}
    \caption{Expected (left) and measured (right) probability distribution of a target Haar-random unitary matrix. The measured matrix achieves a similarity of 95.12\% with respect to the target.}
    \label{Fig:Haar_Random}
\end{figure}

We perform this measurement using classical light, by sequentially injecting coherent light at a wavelength of 1545 nm from a continuous-wave tunable C-band laser source (EXFO T200S-CL) into each input port of the interferometer using polarization-maintaining fibers and a 1$\times$32 optical fiber switch (Santec OSX-100). For each input port, the output power $P_{out, k}$ at all output ports $k$ is recorded using a multi-channel photodiode array (Santec MTA-100), allowing reconstruction of the scattering probabilities $p_{i \rightarrow j}^{(exp)} = \frac{P_{out, j}}{\sum_k P_{out, k}}$. 
An example of target and experimentally measured scattering probabilities for a randomly chosen Haar-random unitary transformation is shown in Figure~\ref{Fig:Haar_Random}. Across ten implemented 12-mode Haar-random unitary transformations, the interferometer achieves an average similarity of $0.94 \pm 0.02$.

\begin{figure}[!ht]
    \centering
    \includegraphics[width = 0.6\textwidth]{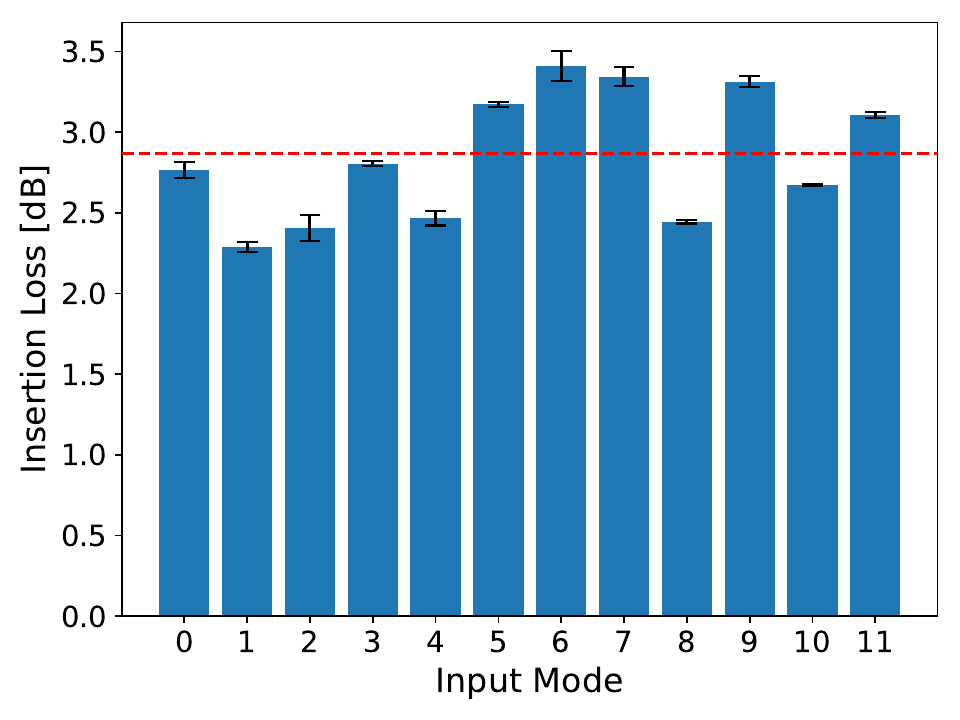}
    \caption{Measured insertion losses of the 12-mode interferometer across all input modes. The average loss of 2.87\,dB is shown as a dashed red line. Uncertainties were taken as the standard deviation of a set of repeated measurements.}
    \label{Fig:Supp_Loss}
\end{figure}

Using the same setup we further characterize the insertion losses across all modes of the interferometer. First, the reference power \(P_{\mathrm{ref},i}\) for each input mode \(i\) is recorded by directly connecting each output port of the fiber switch to the corresponding input port of the photodiode array. The input and output modes of the interferometer are then connected to the fiber switch and the photodiode array, respectively. With the interferometer programmed to implement the identity transformation, the output intensities \(P_{\mathrm{out}, k}\) are measured across all output modes \(k\), for each input mode \(i\). The insertion loss, in dB, for each mode is found to be:
\begin{equation}
 IL_{i} = -10 \cdot \log_{10} \left( \frac{\sum_{k=1}^{12} P_{\mathrm{out}, k}}{P_{\mathrm{ref},i}} \right).   
\end{equation}
The integrated interferometer exhibits average insertion losses of $2.87 \pm 0.37$\,dB and with all insertion losses below $3.41 \pm 0.11$\,dB, as shown in Figure~\ref{Fig:Supp_Loss}.

\subsubsection{Implemented Haar random matrix for measurement runs}
\label{subsec:NetworkMatrix}

Due to limitations in the available equipment, only 8 modes of the installed 12-mode interferometer were utilized for the full measurement runs. The remaining 4 input modes were programmed to route all input power to their corresponding output modes, i.e. identity operation for those 4 modes. 
For the 8 modes that were used, we selected a unitary matrix from a Haar-random distribution, presented in Equation \ref{eq:targetmatrix}.
\begin{equation}
\label{eq:targetmatrix}
\resizebox{\textwidth}{!}{$
\boldsymbol U =
\begin{pmatrix}
    0.47-0.07i & -0.20+0.18i & -0.49-0.51i & -0.22-0.03i & 0.07+0.12i & -0.12+0.04i & 0.06-0.02i & -0.34-0.05i \\
    -0.28+0.21i & -0.41+0.26i & 0.12+0.37i & -0.06+0.13i & -0.07+0.24i & -0.22+0.15i & 0.06-0.04i & -0.55-0.18i \\
    -0.10-0.40i & 0.28+0.01i & 0.20-0.23i & 0.43+0.18i & 0.23+0.10i & -0.56-0.05i & -0.04-0.09i & -0.13-0.20i \\
    -0.14-0.26i & -0.12-0.07i & 0.29-0.14i & -0.57+0.02i & -0.05+0.01i & -0.38+0.17i & 0.27+0.24i & 0.17+0.36i \\
    0.09-0.18i & -0.30+0.39i & 0.16+0.00i & 0.04+0.03i & 0.26-0.53i & 0.08+0.17i & -0.21+0.41i & 0.12-0.27i \\
    0.52+0.06i & 0.29-0.22i & 0.07+0.29i & -0.03+0.08i & -0.22+0.07i & -0.14+0.57i & 0.03+0.15i & 0.03-0.27i \\
    -0.13+0.24i & 0.17-0.24i & -0.17-0.03i & 0.01+0.32i & 0.34+0.21i & 0.10-0.14i & 0.08+0.71i & -0.13+0.00i \\
    0.05+0.09i & 0.03-0.37i & 0.09-0.08i & -0.19+0.48i & -0.11-0.53i & -0.02-0.10i & -0.28-0.20i & -0.36+0.12i
\end{pmatrix}
$}
\end{equation}

\begin{figure}[!ht]
    \centering
	\includegraphics[width=0.8\textwidth]{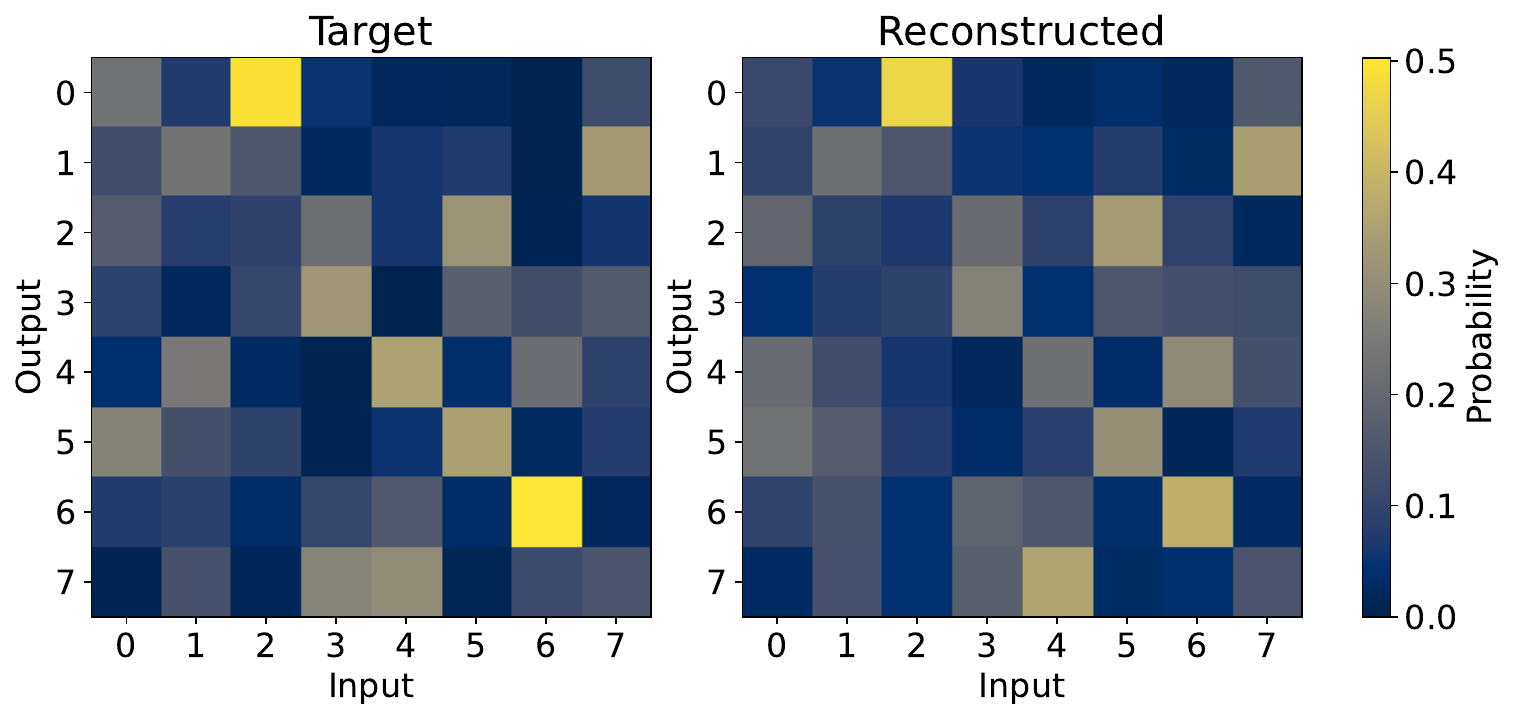}
	\caption{Haar-random matrix implemented for the full measurement runs.
		Implemented target matrix (left) and experimentally reconstructed matrix (right). The reconstructed matrix is obtained from scattershot boson sampling as described in Appendix~\ref{App:expdesign}. The similarity between these two matrices is $97.5\%$.
	}\label{fig:Pmatrix}
\end{figure}

To verify that we implement the desired matrix, we reconstruct the scattering probabilities $p_{i \rightarrow j}^{(exp)}$ using data from the scattershot boson sampling run at $r = 0.176\pm0.002$. 
This is done by heralding on single-photon input-states in input/heralding mode $i$ and count how often the photon is detected in output mode $k$. Similar to the approach with coherent light inputs, the scattering probabilities can then be calculated as $p_{i \rightarrow j}^{(exp)} = \frac{n_{out, j}}{\sum_k n_{out, k}}$ where $n_{out, k}$ are the number of occurrences where the photon was measured in mode $k$.
Evaluating this for all input and output modes, one can use Equation~\ref{eq:similarity} to quantify the similarity. The target and experimentally measured scattering probabilities for the implemented 8-mode matrix are shown in Figure~\ref{fig:Pmatrix} and result in a similarity of $97.5\%$.

\subsection{Photon-Number Resolved Detection}

Photon detection is implemented using a 16-channel superconducting nanowire single-photon detector (SNSPD) system (Single Quantum). 
The SNSPDs have an efficiency $\geq 90\,\%$ and timing jitter $< 20$\,ps. 
The dead time of the detectors is $< 80$\,ns.

We utilize the low timing jitter of our SNSPDs together with our high timing precision time tagger ($< 2$\,ps) (Swabian Instruments Time Tagger X) to achieve intrinsic photon-number resolution up to a maximum of three photons \cite{Schapeler2024,Sidorova2025}. 
In contrast to previously realized pseudo photon-number resolved detection schemes \cite{Deng2023}, the intrinsic photon-number resolution used here gives shot-to-shot photon-number measurements and is enabled by the optimized $\approx 2$\,ps squeezed-light pulses generated in our single-pass waveguide source. 
Sampling experiments are typically operated at average photon numbers of approximately $\langle \bar n \rangle\leq1$ and therefore this photon-number resolution is generally sufficient.

\subsubsection{Heralding detection}

The heralding detection scheme placed at one output of the PBS behind the waveguide is used to herald the number of photons present in each pulse of the pulse train that exits this port of the PBS. This heralding mechanism serves many useful purposes. It can be used to herald the number of photons entering each mode of the interferometer in the SBS configuration as described in Appendix\ref{App:Review} and the photon-number measurements here can be compared to those measured in the signal detection scheme, at the output of the interferometer, to characterize the correlation present in these two signals. In contrast to previous SBS implementations \cite{Bentivegna2015} this PNR detection scheme allows us to herald Fock states with greater than a single photon in any given mode. This is the first experimental realization of such a higher-order SBS setup.

\begin{figure}[!ht]
    \centering
\includegraphics[width=0.7\textwidth]{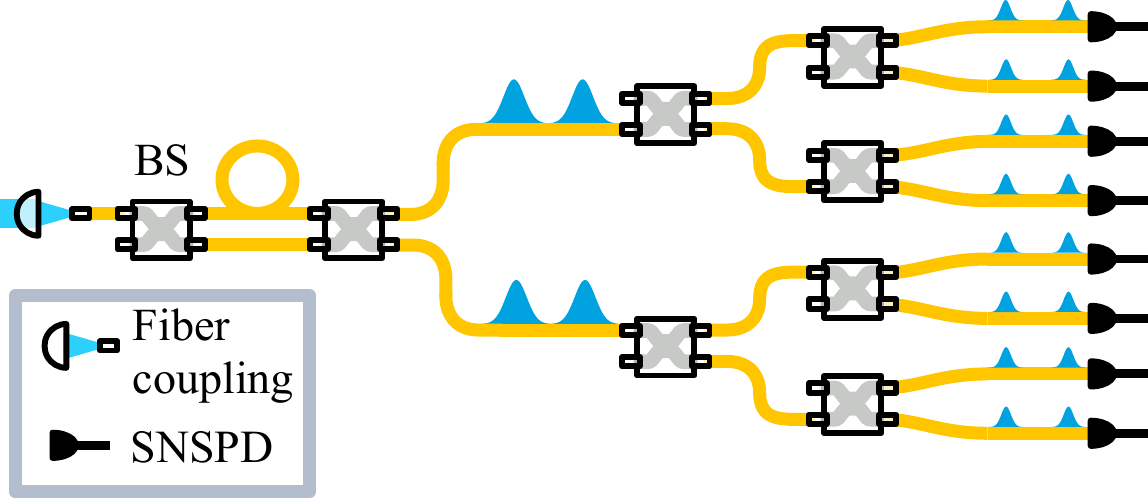}
\caption{Schematic of the multiplexing scheme used for heralding. Input pulses are fiber coupled and temporally and spatially multiplexed using balanced fiber beam splitters (BS) and a fiber delay line. Blue pulses illustrate the bins over which input photons may be found at each step. The light is detected via superconducting nano-wire single photon detectors (SNSPD).}
\label{fig:setup_herald}
\end{figure}

The experimental setup for this demultiplexing scheme is shown in Figure~\ref{fig:setup_herald}. 
We implement 16 detection bins: 8 spatial and 2 temporal modes. The temporal multiplexing is designed to provide a delay of $\tau_d>180$\,ns allowing the detectors to fully recover from detection in the first time bin before the pulse train of the second time bin arrives. 
In addition to the spatial and temporal multiplexing, the heralding also implements the aforementioned intrinsic photon-number resolution for all of the 16 detection bins, mitigating the effect of multiple photons hitting one detector in a single pulse, due to probabilistic splitting of the multiplexing.

Due to the short pulse separation of $12.5$\,ns, which is much faster than the 80\,ns dead time of a single detector, it is necessary to implement a demultiplexing scheme to ensure reliable reconstruction of the input photon number. 
The relatively long dead time of the detector means that detection of photons in any pulse within a single sampling run (typically 8 pulses separated by 12.5\,ns) would blind the detector to the presence of any photons in subsequent pulses within this run. 
The utilized multiplexing scheme, consisting of both spatial and temporal multiplexing, reduces the probability of a second-photon hitting an individual SNSPD that has already detected a photon in a previous pulse, thereby providing a pseudo PNR detection. 

We investigate the effect of detector blinding in our heralding scheme, by calculating the probability of not measuring $n$ photons when $n$ photons where input, i.e. the probability of photon-number misassignment, and the probability of detecting $m\leq n$ photons when $n$ photons where input. 
We do this by tracking the probability of detecting $n$ photons in each pulse of the pulse train and, therefore, how many detectors were set in the blind state in each pulse. 
This is repeated for all possibilities of $n$ photons hitting the detectors at each pulse from the pulse train. 
These combinations are weighted by the possibility of the photon numbers occurring. Finally, we also include the possibility of photons being lost before detection and therefore not blinding the detectors. 
We can show that the main impact of wrong assignments in our heralding setup is caused by photon loss and not detector blinding. 

To simplify the treatment and to account for potentially higher dead times of the detectors, we assume that if a detector has detected a photon and is therefore blinded, it stays blinded for the rest of the 8-pulse pulse train. 
Even though a detector might be partially recovered from a detection in one of the first pulses from the pulse train to one of the last, this assumption gives a lower bound for the detection probabilities. 
Additionally, we treat the detectors as click detectors with zero dark counts. 
This ignores the employed intrinsic photon-number resolution. 
Including the intrinsic PNR would slightly increase the detection efficiency and lower the error probability, but the effect of detector blinding would stay unchanged. 
Neglecting the dark counts is well justified for our SNSPDs that possess dark count rates $<100$ counts per second and are further suppressed via gating.

The three contributions to the detector blinding probability are now discussed in more detail. 
The probability of blinding $k$ from $d$ available detectors with $n$ randomly distributed photons is given by
\begin{equation}
    P_{\mathrm{blind}}(k|n,d) = \binom{d}{k} S(n,k) \frac{k!}{d^n}.
\end{equation}
Where $S(n,k)$ is the Stirling number of the second kind. 
This includes the cases where several photons hit a single click detector.

The probability of a certain photon number $n$ emitted by the source into the heralding arm of our SBS experiment is determined by a thermal photon number distribution with a given mean photon-number $\langle n \rangle$
\begin{equation}
    P_{\mathrm{thermal}}(n) = \frac{\langle n \rangle^n}{(1+\langle n \rangle)^{1+n}}.
\end{equation}
This is because we use a TMSVS as our resource for SBS. 

The experimental detection efficiency $\eta$ of the heralding arm is known from measurement. 
With this we get the probability of having $l$ detection events given $k$ detectors are illuminated by photons
\begin{equation}
    P_{\mathrm{loss}}(l|k) = \binom{k}{l} \eta^l \cdot (1-\eta)^{k-l}.
\end{equation}
With this we are able to calculate the probability of heralding a certain photon-number given an input photon-number for each of the consecutive pulses from our source. 

\begin{figure}[!ht]
\includegraphics[width=0.9\linewidth]{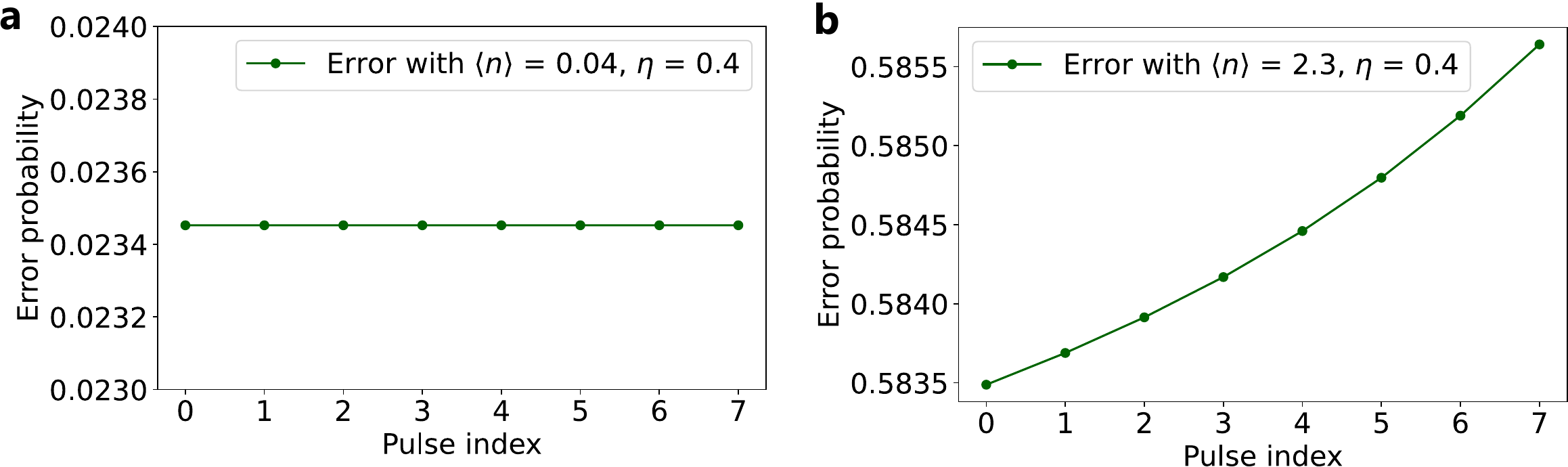}
\caption{Assignment error probability for heralding setup. \textbf{a,} Mean photon-number 0.04, efficiency 0.4. \textbf{b,} Mean photon-number 2.3, efficiency 0.4.}
\label{fig:error_prob}
\end{figure}

The resulting error probability, i.e. the probability of not detecting the exact number of photons present at the input, is shown in Figure~\ref{fig:error_prob}. 
Here we show two different mean photon numbers of 0.04 and 2.3 corresponding to the squeezing parameters 0.2 and 1.2, respectively, for a detection efficiency of 0.4. 
In Figure~\ref{fig:error_prob} \textbf{a} no change is apparent for increasing pulse index, where a pulse index of 0 corresponds to the first pulse of the pulse train and 7 to the last. 
Only for higher pump strength a clear trend of higher errors for increasing pulse index emerges, due to the detector blinding (see Figure~\ref{fig:error_prob} \textbf{b}). 
However, it is apparent that the change in error probability is rather small compared to the offset of the error caused by the detection efficiency of 0.4. 
Note that the error probability has a maximal value of 0.038 and 0.697 for \textbf{a} and \textbf{b}, respectively. 
This is because the input photon-number of 0 photons is always assigned correctly. 

\begin{figure}[!ht]
\centering
\includegraphics[width=0.4\linewidth]{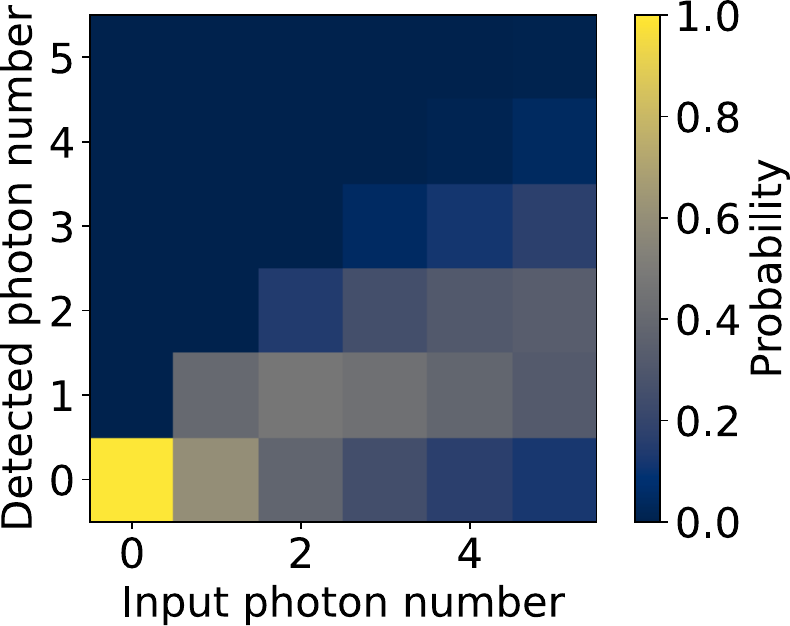}
\caption{Probability of detecting $m$ photons for $n$ input photons. All columns normalized to one.}
\label{fig:assignment_prob}
\end{figure}

To further illustrate that the losses are the main contribution to incorrect assignments of photon numbers we show in Figure~\ref{fig:assignment_prob} the probability of detecting $m\leq n$ photons if $n$ photons are input to the detection. 
In this plot the assignment probability of all 8 pulses from the pulse train are averaged and each input column is normalized to one. The effect of losses is most apparent for 1 photon input. 
The detection probability of one photon is equal to the efficiency of the system $P(1|1) =\eta = 0.4$. 
For higher input photon numbers, the link to the efficiency is not that visually apparent and slightly exaggerated by detector blinding, but the effect of loss is still dominant.

\subsection{System Synchronization}

Due to the complexity of our multi-component setup, precise synchronization is critical. This is achieved by referencing all electronic devices to the mode-locked laser, the repetition rate of which is phase-locked at 80~MHz to an internal RF oscillator, providing a stable master clock. 
This 80~MHz reference is supplied to the time-to-space demultiplexer control box, which generates integer fractions of the clock to drive each EOM in the tree (see section \ref{subsec:demux}).

To reduce the repetition rate of the pump pulses and enable precise temporal control, a digital delay generator (Stanford Research Systems DG645) — synchronized to the 80 MHz master clock using a 1~MHz TTL signal provided by the Menlo Smartcomb unit — drives the pulse picker (EOM1).
It selectively gates the desired number of optical pulses within a defined time window, effectively lowering the repetition rate to accommodate detector dead time while maintaining synchronization across the entire system.
A portion of the pump beam after the pulse picker is recorded and converted to a low-noise electrical signal via a high-speed photodiode, providing the time-tagger reference essential for photon-number-resolving detection \cite{Schapeler2024}. 

For switching between GBS and SBS/TBS, a function generator (Siglent SDG 1032X) generates a 25\,MHz square wave (4\,V amplitude) to drive EOM2. 
A second output of this function generator creates a pulsed signal (200\,$\upmu$s width, 25~MHz repetition rate) that is synchronized to the square wave signal. This signal is sent to the time tagger as a reference to retrieve information about the system configuration (GBS or SBS/TBS configurations).

\subsection{System Characterization}
\label{subsec:syst_char}

\subsubsection{Total Transmission}
\label{subsec:syst_char_tot}

The total transmission of the system, or equivalently system efficiency, is measured using the Klyshko efficiency \cite{Klyshko1980}. 
The system is configured to implement SBS and the identity operation is programmed in the interferometer, i.e. each input mode is routed to the corresponding output mode. 

\begin{figure}[!ht]
        \centering
        \includegraphics[width=0.7\textwidth]{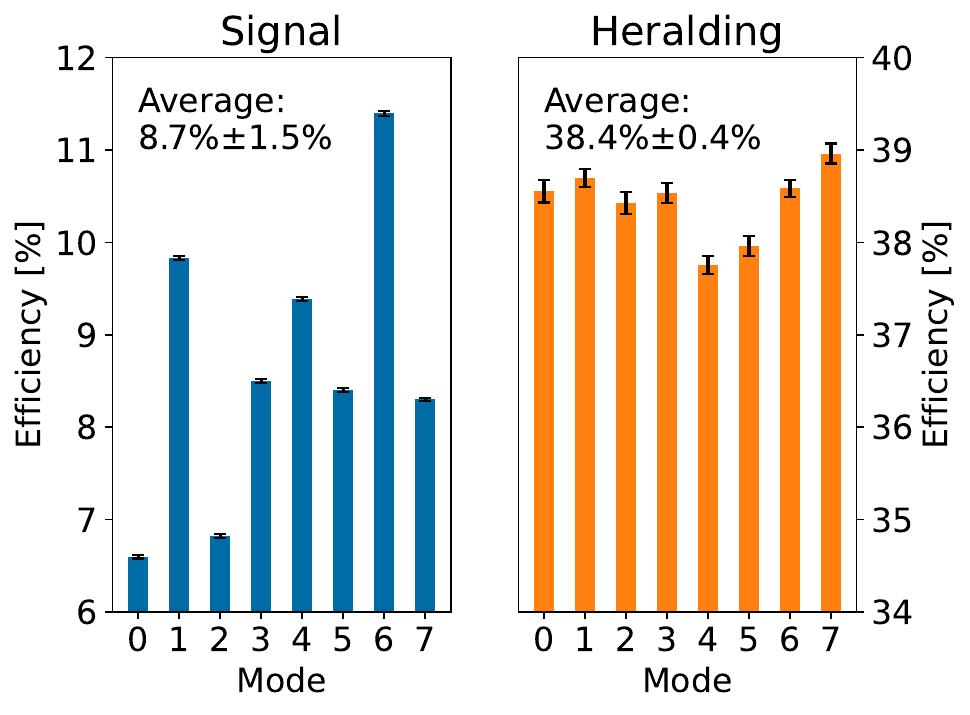}
	\caption{System Efficiencies. The measured Klyshko efficiencies (total system efficiencies) for all eight signal and heralding modes. Uncertainties are given by  counting statistics
	}\label{fig:system_performance}
\end{figure}

The single count rates $H_{i}$ and $S_{i}$ for the heralding and signal modes, respectively, and the coincidences $C_{i}$ between these two modes modes $i = 0, 1, \dots, 7$ are collected. 
The efficiencies can then be calculated using $\eta^{(i)}_{\mathrm{herald}} = \frac{C_{i}}{S_{i}}$ and $\eta^{(i)}_{\mathrm{signal}} = \frac{C_{i}}{H_{i}}$. 
The efficiencies measured in this way take into account all sources of loss such as waveguide loss, transmission and detection inefficiencies. The measured efficiencies can be seen in figure \ref{fig:system_performance} and show similar efficiencies for all eight heralding modes with an average of $38.4\,\%\pm0.4\,\%$, and signal path efficiencies between $6.6\,\%\pm0.1\,\%$ and $11.4\,\%\pm0.1\,\%$, with an average efficiency of $8.7\,\%\pm1.5\,\%$.

\subsubsection{Temporal Overlap}
To ensure indistinguishability of the modes within the interferometer and achieve high-visibility interference, precise temporal overlap is essential.
This is accomplished using the PLCUs detailed in Section \ref{subsec:PLCS}. 
Due to the temporal multiplexing of the squeezed-light source, each input mode propagates through a distinct fiber length.
These path lengths must be controlled to within a fraction of the pulse duration (typically less than 0.1 ps for our system) to maintain temporal overlap.

To characterize the performance of our path length compensation, we perform heralded signal-signal HOM measurements \cite{ou1999photon,mosley2008conditional,jin2015spectrally}.
This is achieved by implementing 50:50 beam splitters between pairs of modes under test and using the heralding arm to flag single-photon Fock states in both input modes simultaneously—creating a standard HOM interference configuration, where perfectly indistinguishable photons are expected to bunch perfectly (zero coincidences).
Imperfect temporal overlap introduces distinguishability - thereby increasing the coincidence rates. By tuning the temperature of the PLCU units, we vary the temporal delay and record the number of coincidences from which the visibility of the HOM interference can be determined. 
The optimization is performed relative to a common reference (mode 7), and the measurement is parallelized across all eight input pulses using up to four beam splitters, enabling simultaneous characterization of four HOM dips and ensuring uniform temporal alignment across the entire interferometer.

\begin{figure}[!ht]
        \includegraphics[width=0.49\textwidth]{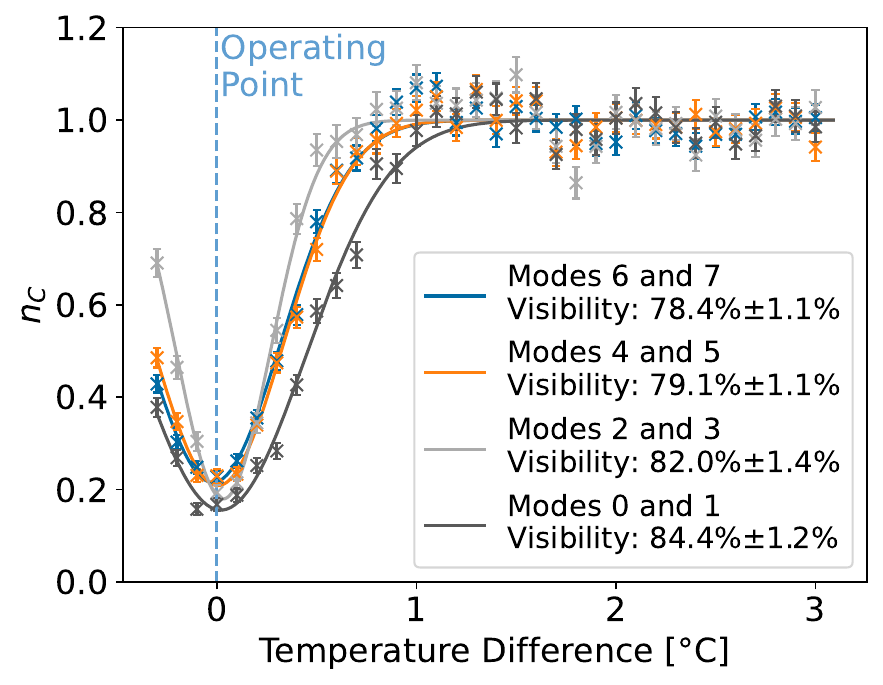}
	\hfill
	\includegraphics[width=0.49\textwidth]{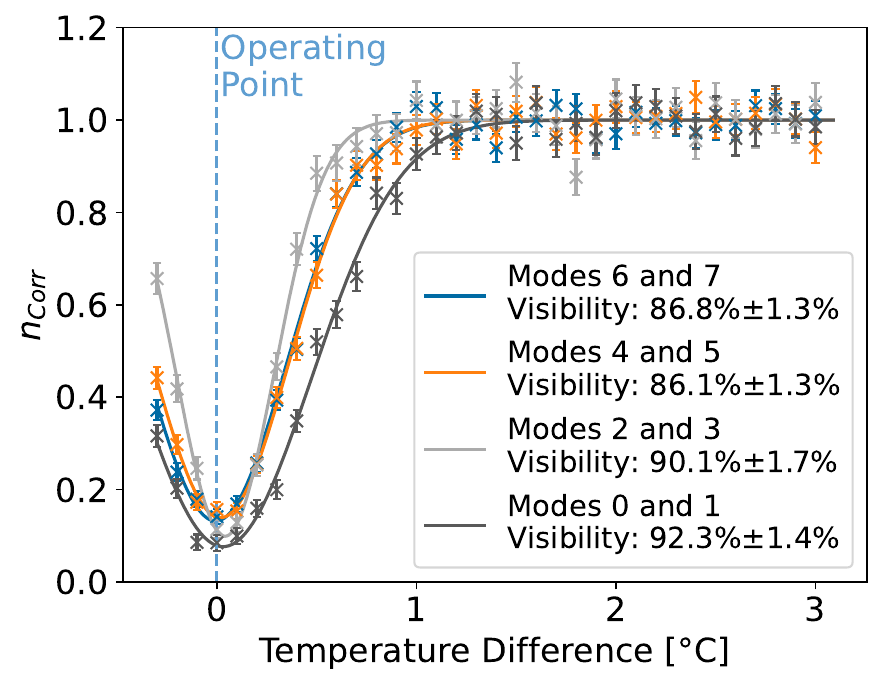}
	\caption{Signal-signal HOM dips for different combinations of modes. Left: Before correcting for "false" coincidence counts and normalizing by mean photon-number. Right: After correction and normalization. The ``Operating Point" (blue dashed line at zero delay) indicates the temperature setpoint of corresponding modes after delay optimization. Uncertainties are given by counting statistics.}
    \label{fig:signalsignalHOM}
\end{figure}

Figure \ref{fig:signalsignalHOM}(a) displays four simultaneously measured heralded signal-signal HOM dips taken at a detected mean photon number of $\approx 0.003$, leading to visibilities in the range from $78.4\,\%\pm1.1\,\%$ to $84.4\,\%\pm1.2\,\%$ 
across different input channel pairs. 
These values are limited by any source of distinguishability in the photons and contributions due to higher-order photon number events. The maximum achievable HOM visibility is limited by the effective spectro-temporal mode number $K$ of the source (see Section \ref{subsec:second-order}). 
Given that $V_{\mathrm{max}} = 1/K$ and the measured second-order auto-correlation of the source of $g^{(2)} = 1.95\pm0.03$ we would expect the visibility to obtain a maximum value of $V_{\mathrm{max}} = 95.2\,\%\pm2.8\,\%$. 

The observed heralded HOM visibilities are seen to lie significantly below the theoretical maximum. Sources of this reduction could be multi-photon contamination from higher-order PDC events and pump power drifts.
The former contaminates the measurement with unwanted/false coincidence counts that will reduce the visibility. Power drifts can lead to both, a decrease or increase of the visibility, depending on wether the pump power inside the dip is lower (increased visibility) or higher (reduced visibility) than outside of the dip. We apply a correction protocol to negate the impact of higher photon-number contamination using a method described in the following subsection.
This protocol yields corrected visibilities between $86.1\,\%\pm1.3\,\%$ and $92.3\,\%\pm1.4\,\%$ , as shown in Figure \ref{fig:signalsignalHOM}(b).  
It is likely that the residual gap stems from imperfect beam splitter ratios, but requires further investigation.

\begin{figure}[!ht]
        \centering
        \includegraphics[width=0.5\textwidth]{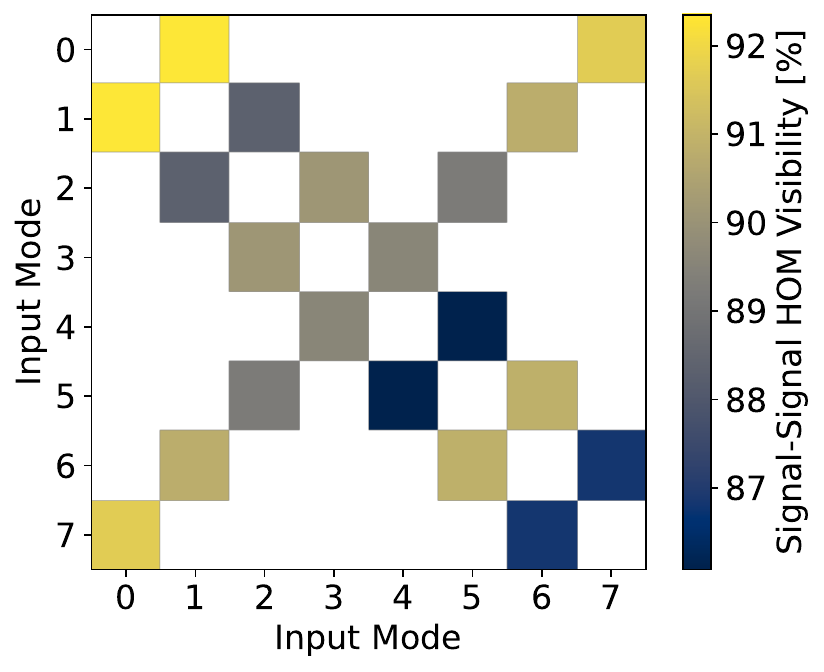}
	\caption{Overview of all measured signal-signal HOM visibilities. The average visibility is $89.7\,\%\pm2.0\,\%$. Note that the upper right triangle shows the same data as the lower left triangle.
	}\label{fig:HOM_overview}
\end{figure}

To extend the interference characterization we repeat the HOM interference measurement for other input mode combinations, including modes 0 and 7—which exhibit the largest fiber length disparity. This particular combination allows to assess the impact of path-length differences and fiber dispersion on the interference visibility. After correcting for multi-photon events and pump drift, these two modes yielded a visibility of $91.7\,\%\pm2.1\,\%$, consistent with the theoretical maximum within error margins, confirming that chromatic dispersion in the fiber links does not limit interference quality. As shown in Figure \ref{fig:HOM_overview}, the average corrected visibility across all measured mode pairs is $89.7\,\%\pm2.0\,\%$.
The residual deviations are primarily  attributed to the non-ideal 50:50 beam splitters implemented in the integrated interferometer, which introduce distinguishability and reduce visibility.

\subsubsection{Multi-photon correction}
HOM measurements typically assume single-photon input states and, therefore, untrustworthy results are obtained when this condition no longer holds. Owing to the nature of PDC sources, this condition will break down as the mean photon number of the generated state increases, the impact of which can be accounted for with a more detailed treatment, as we detail here.

Owing to the fact that we have intrinsic PNR in our detectors, we are able to filter our measured four-fold coincidence counts  $n_{\textrm{C}}$ for events where only one photon is detected in all four detectors (both heralding channels and both beam splitter outputs). 
This aids in removing unwanted events, but there is still a chance that a higher photon number event registers as a true four-fold single-photon event due to optical loss. We label these events as ``false'' coincidences $n_{\textrm{F}}$. 
 
To account for false coincidences arising from multi-pair events, we consider the most likely cases involving an additional photon pair: (i) two photon pairs are generated in input (1) and one pair in input (2), and (ii) one pair in input (1) and two pairs in input (2). 
For case (i), incorporating individual efficiencies, $\eta_{\mathrm{herald}}^{(1)}$, $\eta_{\mathrm{herald}}^{(2)}$, $\eta_{\mathrm{signal}}^{(1)}$ and $\eta_{\mathrm{signal}}^{(2)}$, the false coincidence rate is:
\begin{equation}
    n_\textrm{F}^{(2, 1)} \approx \frac{1}{2} \left(1 - \eta_{\mathrm{herald}}^{(1)}\right)\left(1 - \eta_{\mathrm{signal}}^{(2)}\right)\eta_{\mathrm{herald}}^{(1)}\eta_{\mathrm{herald}}^{(2)}\eta_{\mathrm{signal}}^{(1)}\eta_{\mathrm{signal}}^{(1)} \langle n^{(1)} \rangle^2 \langle n^{(2)} \rangle n_\textrm{T},
\end{equation}
where $\langle n^{(1)} \rangle^2 \langle n^{(2)} \rangle \approx p(2, 1)$ is the probability of the pair configuration, and $n_\textrm{T}$ is the total number of measurement triggers. 
Assuming symmetric heralding efficiencies ($\eta_{\mathrm{herald}}^{(1)} = \eta_{\mathrm{herald}}^{(2)} = \eta_{\mathrm{herald}}$) and mean photon numbers ( $\langle n^{(1)} \rangle = \langle n^{(2)} \rangle = \langle n \rangle$), and defining the signal efficiency as the average $\eta_{\mathrm{signal}} = \frac{1}{2}(\eta_{\mathrm{signal}}^{(1)} + \eta_{\mathrm{signal}}^{(2)})$ to account for unknown loss distribution before/after the beam splitter, the total false coincidences (summing both cases) simplifies to:
\begin{equation}
    n_\textrm{F} = n_\textrm{F}^{(2, 1)} + n_\textrm{F}^{(1, 2)} \approx (1 - \eta_{\mathrm{herald}})(1-\eta_{\mathrm{signal}})\eta_{\mathrm{herald}}^2 \eta_{\mathrm{signal}}^2 \langle n \rangle^3 n_\textrm{T}.
\end{equation}
The factor $\frac{1}{2}$ is absorbed by including both configurations equally. 
We can then obtain the corrected number of coincidences by removing these counts and normalizing by the mean photon numbers in the heralding detection in order to account for power drifts, yielding
\begin{equation}
    n_{\mathrm{Corr}} \approx \frac{n_\textrm{C} - n_\textrm{F}}{\langle n \rangle^2_{\mathrm{herald}}}.
\end{equation}
The impact of this correction for the measured HOM dips is shown in Figure \ref{fig:signalsignalHOM}.

\section{Nonclassicality Criterion}
\label{App:theory}

The matrix of moments (MoM) of normally ordered photon-number correlations from Refs. \cite{Agarwal1992,SRV05} is the basis of our nonclassicality criterion.
The $M$-mode, second-order MoM takes the form
\begin{equation}
    \mathrm{MoM}=\begin{bmatrix}
        1 & \langle{:} \hat n_1 {:} \rangle & \hdots & \langle{:} \hat n_M {:} \rangle
        \\
        \langle{:} \hat n_1 {:} \rangle & \langle{:} \hat n_1\hat n_1 {:} \rangle & \hdots & \langle{:} \hat n_1\hat n_M {:} \rangle
        \\
        \vdots & \vdots & \ddots & \vdots
        \\
        \langle{:} \hat n_M {:} \rangle & \langle{:} \hat n_M\hat n_1 {:} \rangle & \hdots & \langle{:} \hat n_M\hat n_M {:} \rangle
    \end{bmatrix},
\end{equation}
where ${:}\cdots{:}$ denotes the normal-ordering prescription.
It has been shown that the matrix of moments (MoM)  is positive semidefinite -- i.e., has only non-negative eigenvalues -- for classical light, $\mathrm{MoM}\geq0$.
Thus, if the smallest eigenvalue of MoM is negative, nonclassicality is verified, $\mathrm{MoM}\ngeq 0$.

One can apply the invertible transformation matrix
\begin{equation}
    T=\begin{bmatrix}
        1 & 0 & \hdots & 0
        \\
        -\langle{:} \hat n_1 {:} \rangle & 1 & \hdots & 0
        \\
        \vdots & \vdots & \ddots & \vdots
        \\
        -\langle{:} \hat n_M {:} \rangle & 0 & \hdots & 1
    \end{bmatrix}
\end{equation}
to the MoM,
\begin{equation}
    T\, \mathrm{MoM}\, T^\dag
    =
    \begin{bmatrix}
        1 & 0 & \hdots & 0
        \\
        0 & \langle{:} \hat n_1\hat n_1 {:} \rangle - \langle{:} \hat n_1{:}\rangle\langle{:}\hat n_1 {:} \rangle & \hdots & \langle{:} \hat n_1\hat n_M {:} \rangle - \langle{:} \hat n_1{:}\rangle\langle{:}\hat n_M {:} \rangle
        \\
        \vdots & \vdots & \ddots & \vdots
        \\
        0 & \langle{:} \hat n_M\hat n_1 {:} \rangle - \langle{:} \hat n_M{:}\rangle\langle{:}\hat n_1 {:} \rangle & \hdots & \langle{:} \hat n_M\hat n_M {:} \rangle - \langle{:} \hat n_M{:}\rangle\langle{:}\hat n_M {:} \rangle
    \end{bmatrix}.
\end{equation}
Such a transformation (also known as a conjugation) is known not to alter the signs of the eigenvalues.
Also, the first row and column pertain to the non-negative eigenvalue one and, therefore, can be ignored in the following nonclassicality analysis.

The normal orders of the used expressions in the MoM are well known, ${:} \hat n_j{:}= \hat n_j$ and ${:} \hat n_j\hat n_k{:}=\hat n_j\hat n_k-\delta_{j,k}\hat n_j$, where $\delta_{j,k}=1$ for $j=k$ and zero otherwise.
Thus, the nonclassicality criterion via the transformed MoM can be recast as
\begin{equation}
    T\, \mathrm{MoM}\, T^\dag\cong\boldsymbol{C}-\boldsymbol{B}\ngeq0,
\end{equation}
where "$\cong$" denotes ignoring the first row and column and
\begin{equation}
    \boldsymbol{C}=[\langle \hat n_j\hat n_k\rangle-\langle \hat n_j\rangle\langle\hat n_k\rangle]_{j,k\in\{1,\ldots,M\}}
    \quad\text{and}\quad
    \boldsymbol{B}=\mathrm{diag}[\langle \hat n_j\rangle]_{j\in\{1,\ldots,M\}}
\end{equation}
describing a matrix of (co-)variances $\mathrm{Cov}(n_j,n_k)=\langle \hat n_j\hat n_k\rangle-\langle \hat n_j\rangle\langle\hat n_k\rangle$ and a matrix of bounding values, the mean photon numbers at each output mode, respectively.

\section{Simulations}
\label{App:simulations}

We performed simulations using The Walrus library \cite{Gupt2019} to provide a theoretical context for the initial understanding of the  results presented Fig. 4\textbf{b}. The system configuration differs fundamentally between the GBS and SBS sampling regimes, leading to their distinct features. Consequently, our simulations employed different parameters for each case. To ensure statistical significance, one million samples were generated per squeezing parameter in both sampling regimes.

\subsection{GBS simulations}

\begin{figure}[!ht]
    \centering
    \includegraphics[width = 1\textwidth]{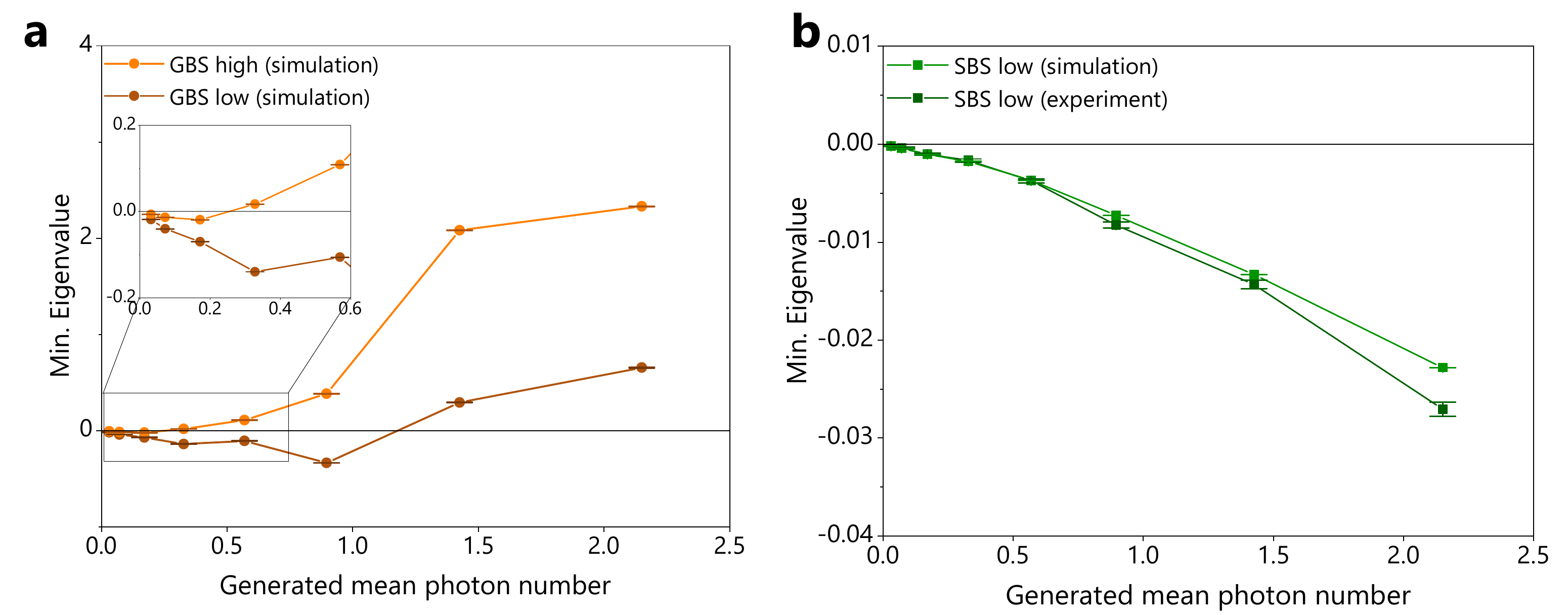}
    \caption{ 
     System simulations in Walrus. \textbf{a,} Simulated GBS performance: The range of minimum eigenvalues (highest: orange; lowest: brown) from 9 lossless simulations with different input phases. 
    \textbf{b,} In contrast to a, the lowest minimum eigenvalues for SBS under experimental loss conditions (simulated: green; experimental data from the Results session: dark green).}
    \label{Fig:GBSandSBSsim}
\end{figure}

To simulate the GBS system, eight SMSV states are generated and propagated through a unitary corresponding to the experimentally implemented 8x8 unitary matrix. The system is modeled with perfect system transmission. Given the known sensitivity of GBS to the input phases of eight SMSV states, the simulation was repeated 9 times, each with a different set of input phases that remain constant throughout each repetition.  All simulations utilized a photon-number cutoff of 15, defined as max\_photons in the Walrus library, to minimize the impact of incorrectly assigning photon number events. For each fixed mean photon number, the highest and lowest minimum eigenvalues obtained from all 9 phase configurations were determined, as illustrated in Figure~\ref{Fig:GBSandSBSsim}\textbf{a}.

The simulated results show strong agreement with the experimental data: at low mean photon numbers both the high and low minimum eigenvalues are negative. Near a mean photon number of 0.5, the lowest minimum eigenvalues begin to increase, while the highest values start to rise already at a mean photon number of 0.33. Notably, these simulations incorporated no loss, a photon-number truncation at 15, and no phase drift during a single simulation run. Consequently,  we have provided strong evidence that the behavior observed experimentally in Fig. 4\textbf{b} of the main text cannot be attributed to these system imperfections.

\subsection{SBS simulations}

To simulate the SBS system, 8 TMSV states were first generated from a PDC process.  One mode of each TMSV state was propagated through the same 8×8 unitary interferometer as in the GBS simulation, while the second mode was detected using a PNR heralding detector. In contrast to the GBS simulations, the SBS model incorporated losses of 0.913 in the interferometer channels and 0.616 in the heralding channels, corresponding to the experimentally determined losses of the system. As was the case for the GBS simulation, we once again employed a photon-number resolution cutoff of 15. The resulting minimum eigenvalues are presented alongside the experimental data in Figure~\ref{Fig:GBSandSBSsim}\textbf{b}.

The simulations qualitatively reproduce the experimental trend of decreasing minimum eigenvalue magnitudes with increasing mean photon number. The observed offset, however, points to unmodeled effects. Potential contributors include the experimental photon-number resolution limit of 3 photons or nonuniform loss between the interferometer channels.

\subsection{Impact of photon-number resolution cutoff}
\label{App:impactpnr}
\begin{figure}[!ht]
        \includegraphics[width=0.49\textwidth]{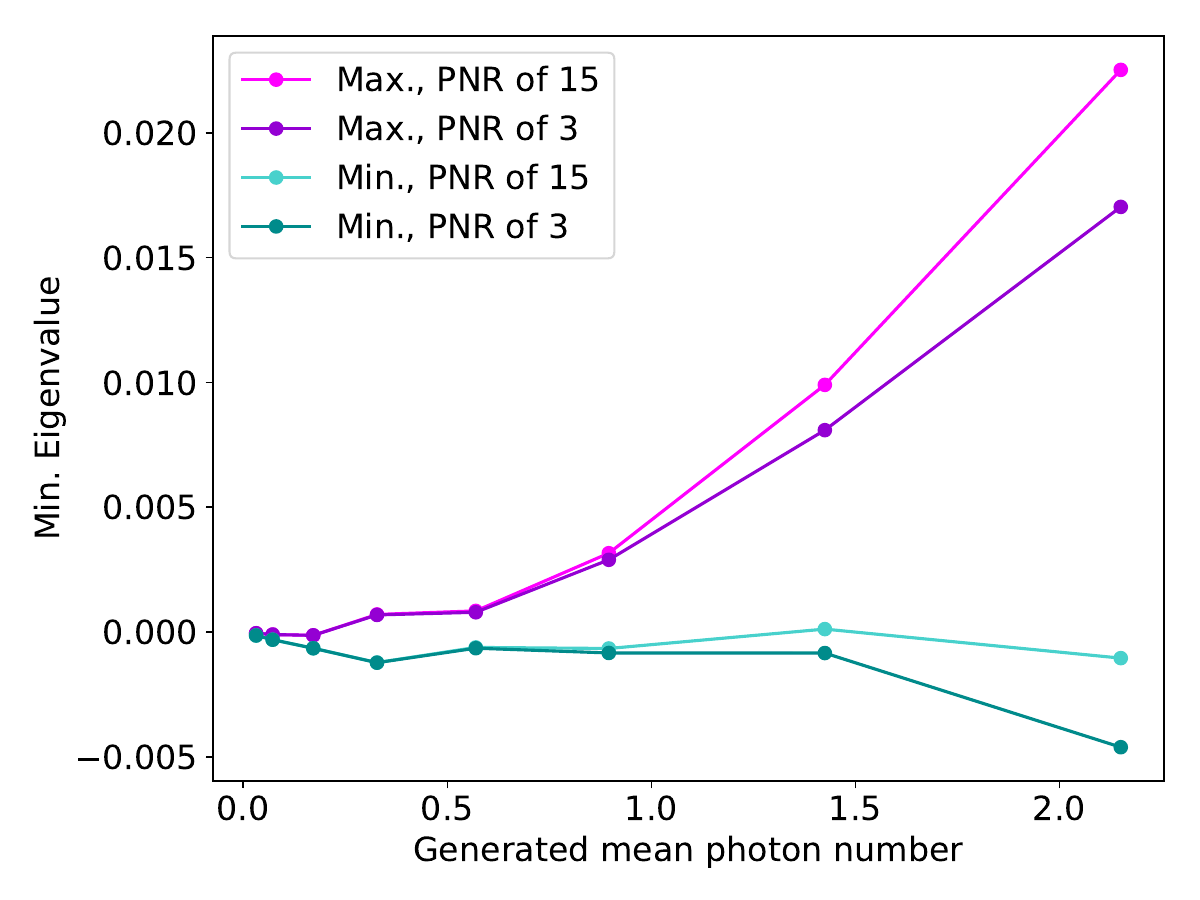}
	\hfill
	\includegraphics[width=0.49\textwidth]{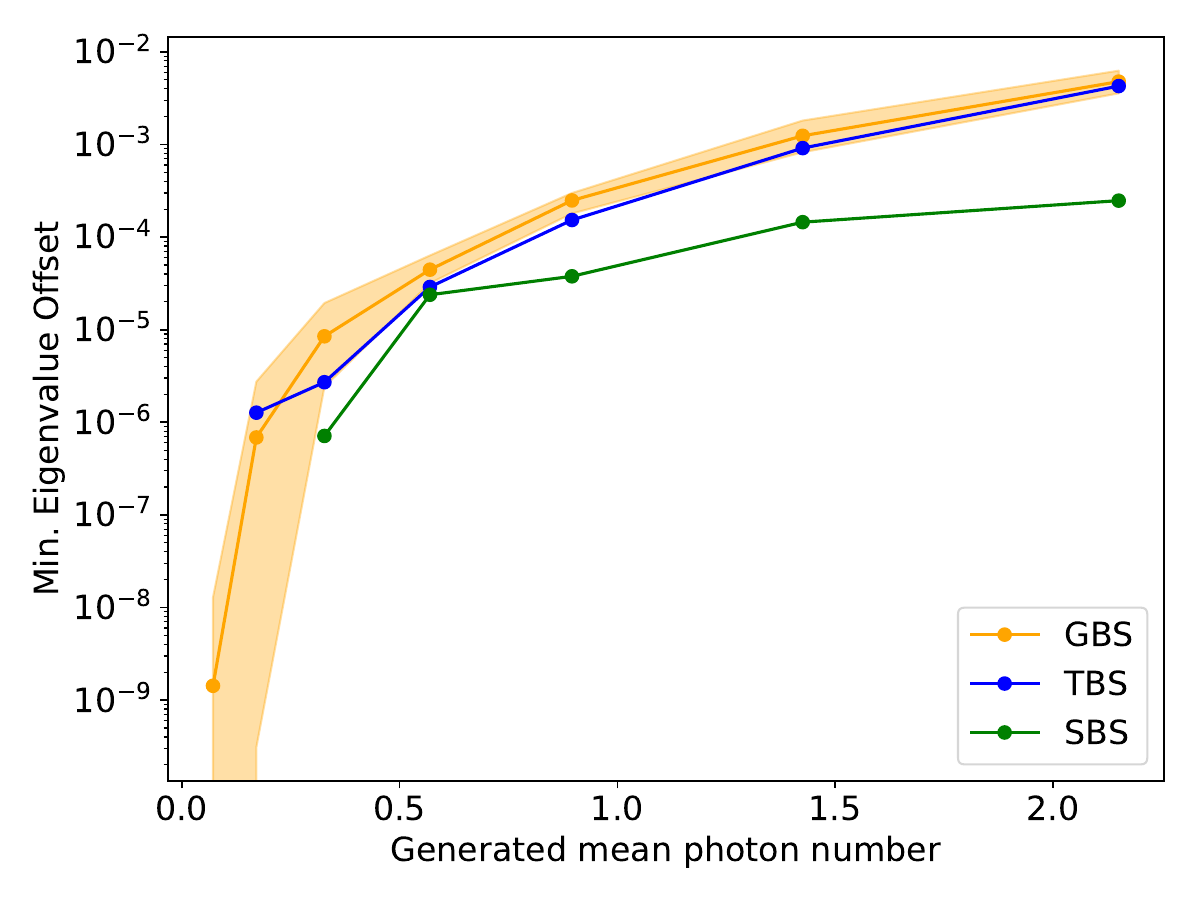}
	\caption{Impact of PNR cutoff on the non-classicality. Left: Minimum and maximum minimum eigenvalues for GBS simulations. The PNR cutoff is seen to artificially decrease the minimum eigenvalue at generated mean photon numbers above 1. Right: For simulated SBS, TBS and GBS datasets, the offset in the minimum eigenvalues due to PNR cutoff, defined as Min. Eig.[PNR=15] - Min. Eig.[PNR=3], is shown. For GBS, the shaded region indicates the range of offsets for different input phase configurations. Data points with an offset of less than $10^{-10}$ are omitted.}
    \label{fig:pnrall}
\end{figure}

To investigate the impact of PNR cutoff (PNR of 3) at the outputs of the interferometer, we perform additional simulations of the GBS and SBS configurations with all parameters set to those matching the physical system, including loss.
In these simulations, we first generate $1$ million samples using a photon-number cutoff of $15$ to produce a dataset with negligible impact due to PNR cutoff. In the case of GBS we sample for $9$ different random input phase configurations. From this dataset, the minimum eigenvalues are determined. Next, all photon numbers exceeding $3$ are reassigned to a photon number of exactly $3$, thereby encapsulating the effect of PNR cutoff in the experiment. The resulting minimum eigenvalues for this dataset are then also calculated and the comparison between the two cases is plotted in Figure \ref{fig:pnrall} (left)
.

The investigation reveals that the impact of PNR cutoff is to artificially decrease the measured minimum eigenvalues. This effect is negligible for the simulated generated mean photon numbers below 0.9. For the three GBS data points with the highest generated mean photon numbers (of $0.896$, $1.424$, and $2.152$), the difference in the minimum eigenvalue due to the PNR cutoff is on the order of the experimental values, as shown in Figure \ref{Fig:PaQSResults} and Table \ref{tab:MainResults}. We note that, even at the highest generated mean photon number of $2.152$, the probability to detect more than $3$ photons in at least one single output channel is less than $0.9$\,\%.

To show the impact of PNR cutoff on the minimum eigenvalues obtained in TBS and SBS configurations, a similar procedure to that described for GBS is applied for simulations of SBS and TBS. Figure \ref{fig:pnrall} (right) shows the offset in minimum eigenvalue that is observed due to the PNR cutoff. TBS is affected in much the same way as GBS, while data from the SBS simulation is significantly less impacted. This is very likely due to the fact that only heralding patterns that appear at least $100$ times are considered  in order to ensure statistical significance, as described in \ref{App:measurement}.

\section{Measurement run overview}
\label{App:measurement}

\subsection{Data Acquisition}

We define a measurement run as the data accumulated over a nearly 12 minute period at a single mean photon number. During a measurement run, the system constantly switches between SBS/TBS and GBS configurations. The system switches between these two configurations every 20\,s and begins recording once the first switch into GBS acquisition mode is signaled. During these 20\,s, data is integrated into separate files every 100\,ms, i.e. files containing 0.1\,M samples due to the 1\,MHz repetition rate of our experiment. Approximately $7,200$ files are generated in this way, resulting in a total measurement time of nearly 12 minutes for each measurement run made at a single mean photon number. Data recorded in a range of $\pm 100$~ms from each switching point is discarded to ensure system stability for all of the analyzed data. This procedure will lead to small differences in the number of files recorded during each switch. In order to ensure statistical significance, heralding patterns that occur less than 100 times are not considered in all of the following analyses.

This procedure is repeated as the mean photon numbers of the generated states is varied to generate the full dataset presented in this work. The mean photon number (or equivalently the squeezing parameter) of the SMSV or TMSV states exiting the waveguide are varied by tuning the pump power driving the process. The targeted squeezing parameters and the implemented squeezing parameters, inferred from other experimental parameters, are given in Table \ref{tab:SqueezingDeciBell}.

\begin{table}[!ht]
	\caption{%
		Squeezing parameters used in experimental runs. We report the targeted (left) and reconstructed (right) squeezing levels and generated mean photon numbers determined from the experimental data.}
        \label{tab:SqueezingDeciBell}
	\begin{tabular}{ll}
    \textbf{Targeted}
		\\
		\hline\hline
		parameter $r$ & $\langle n \rangle_{\mathrm{gen}}$
		\\
		\hline
		0.20 & 0.04
		\\
		0.30 & 0.09
		\\
		0.45 & 0.22
		\\
		0.60 & 0.41
		\\
		0.75 & 0.68
		\\
		0.90 & 1.05
		\\
        1.05 & 1.57
        \\
        1.20 & 2.28
        \\
		\hline\hline
 \end{tabular}
 \quad
 \begin{tabular}{llll}
 \textbf{Reconstructed}
		\\
		\hline\hline
		parameter $r$ & $\Delta r$ & $\langle n \rangle_{\mathrm{gen}}$ & $\Delta\langle n \rangle$
		\\
		\hline
		0.176 & 0.002 & 0.031 & 0.001
		\\
		0.265 & 0.002 & 0.072 & 0.001
		\\
		0.403 & 0.003 & 0.172 & 0.003
		\\
		0.545 & 0.004 & 0.328 & 0.005
		\\
		0.697 & 0.006 & 0.569 & 0.011
		\\
		0.843 & 0.009 & 0.896 & 0.021
		\\
        1.012 & 0.012 & 1.424 & 0.045
        \\
        1.176 & 0.018 & 2.152 & 0.090
        \\
		\hline\hline
 \end{tabular}
\end{table}

To reconstruct the squeezing parameter $r$ and the generated mean photon number $\langle n\rangle _{\mathrm{gen}} $, i.e. before loss, we correct the measured mean photon numbers using the measured Klyshko efficiencies of the heralding channels ${\eta^{(i)}_{\mathrm{herald}}}$ (see section\ref{subsec:syst_char_tot}), 
\begin{equation}
    \langle n \rangle _{\mathrm{gen}}^{(i)} = \frac{\langle n\rangle  ^{(i)}_{\mathrm{meas}} }{\eta^{(i)}_{\mathrm{herald}}},
\end{equation}
where $\langle n\rangle ^{(i)}_{\mathrm{meas}}  $ is the measured mean photon number in channel $i$.
From $\langle n\rangle _{\mathrm{gen}}^{(i)}$, the corresponding (average) squeezing parameter $r$ is obtained using,
\begin{equation}
    r = \frac{1}{8}\sum_i{\mathrm{arcsinh}}\left(\sqrt{\langle n\rangle _{\mathrm{gen}}^{(i)}}\right).
\end{equation}
The corresponding quadrature squeezing level in decibels is then:
\begin{equation}
{\mathrm{Squeezing (dB)}}=20 \cdot r \cdot \mathrm{log}_ 
{10}(e) \approx 8.686 \cdot r.
\end{equation}
This accounts for system loss and enables accurate comparison between targeted and achieved squeezing parameters.

\subsection{Pearson Coefficient}

\begin{figure}[!ht]
    \centering
    \includegraphics[width = \textwidth]{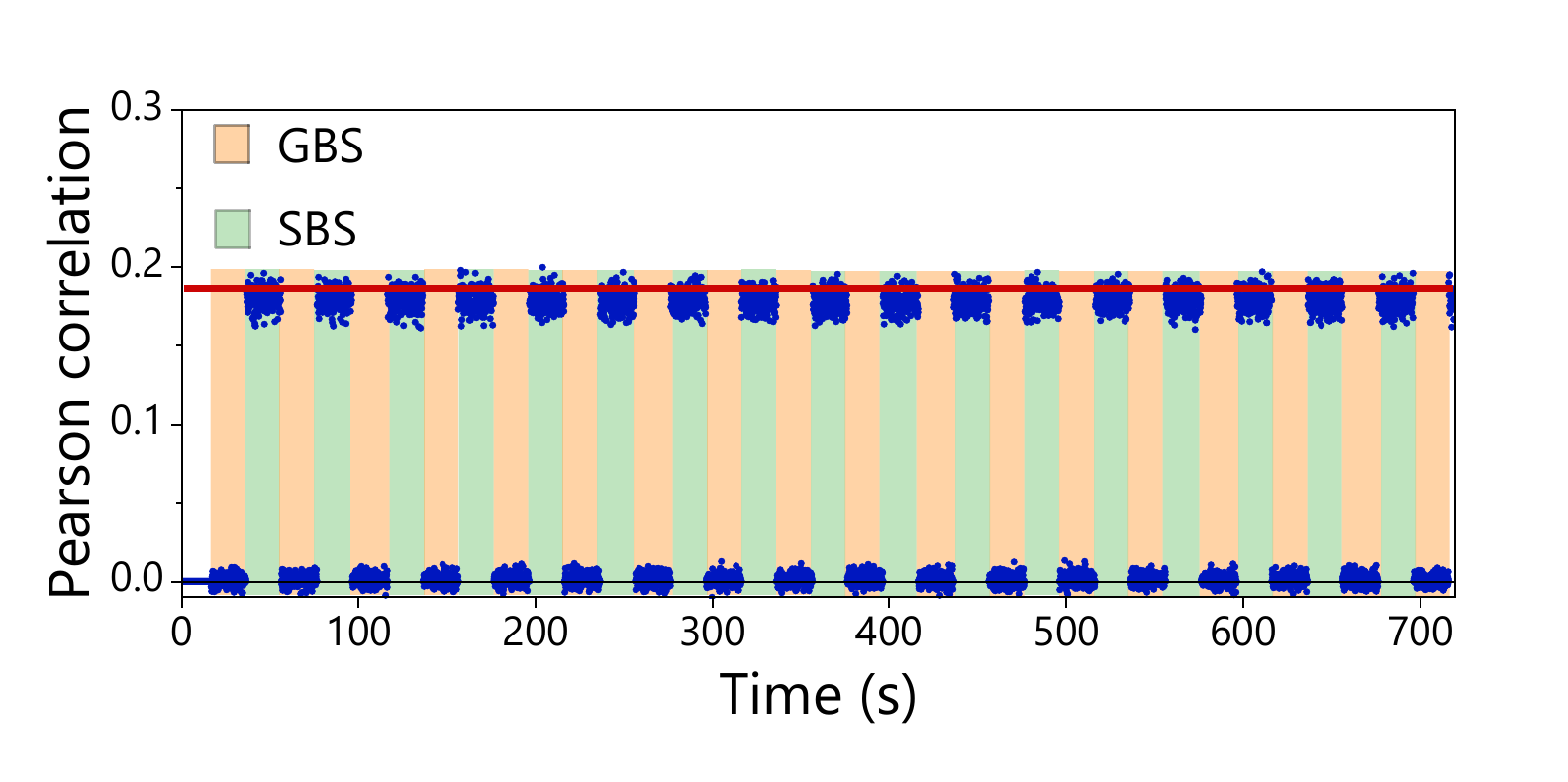}
    \caption{Pearson Coefficient. Measured Pearson coefficient for the $\langle n \rangle = 0.031$ run, with expected values for GBS (black line) and SBS (red line) indicated. The system is switched between these configurations every 20\,s as indicated by the shaded regions.}
    \label{Fig:Pearson}
\end{figure}

Although we switch between producing TMSV and SMSV states, the pump power and following experimental configuration is constant and therefore the total number of photons observed is identical in both the SBS and GBS configurations. Thus, for assessing the correlations between signal and heralding modes, we apply the Pearson correlation coefficient,
 	\begin{equation}
 		\label{eq:Pearson}
 		\gamma=\frac{\mathrm{Cov}(m,n)}{\sqrt{\mathrm{Var}(m)\mathrm{Var}(n)}},
 	\end{equation}
 	where $m=\sum_{j=1}^M m_j$ and $n=\sum_{j=1}^M n_j$.
 	Further recall that $\mathrm{Cov}(m,n)=\mathrm E(mn)-\mathrm E(m)\mathrm E(n)$, $\mathrm{Var}(m)=\mathrm{Cov}(m,m)$, and $\mathrm{Var}(n)=\mathrm{Cov}(n,n)$.
 Ideally, we find the values $1$ and $0$ for perfect correlations and the uncorrelated case, respectively.
 Including losses, we expect that Equation \eqref{eq:Pearson} yields the values
 	\begin{equation}
 		\label{eq:PearsonLoss}
 	\begin{aligned}
 		\gamma_\mathrm{GBS}={}&0
 		\quad\text{and}
 		\\
 		\gamma_\mathrm{SBS}={}&\sqrt{\frac{\eta_\mathrm{herald}\eta_\mathrm{signal}}{
 			[1{-}(1{-}\eta_\mathrm{herald})\tanh^2 r]
 			[1{-}(1{-}\eta_\mathrm{signal})\tanh^2 r]
 		}},
 	\end{aligned}
 	\end{equation}
 assuming TMSV states for the latter coefficient.
 The Pearson coefficient provides an on-the-fly method of ensuring that switching between GBS and SBS/TBS configurations is functioning as intended and also confirms high quality interference at the PBS in the GBS case. See Table \ref{tab:PearsonCoeff} for the corresponding estimates in our experiments. Using the losses determined via our Klyshko measurements allows us to compare the predicted Pearson coefficients to those which are measured, as reported in Table \ref{tab:PearsonCoeff}. We see that although the predicted and measured values differ by less than 10\% across all squeezing parameters, this difference is significantly larger than the expected uncertainty. This discrepancy likely arises due to unaccounted errors when determining the inferred squeezing values and system losses. The measured Pearson correlation coefficient for the data generated at a mean photon number of 0.031 is shown in Figure \ref{Fig:Pearson}.

\begin{table}[!ht]
	\caption{Summary of Pearson correlation coefficients across all measurement runs. The predicted (left) and measured (right)  Pearson coefficients (see Equation \eqref{eq:PearsonLoss}) for all implemented squeezing parameters $r$. Uncertainties are given by counting statistics.
	}\label{tab:PearsonCoeff}
	\begin{tabular}{lll}
    \textbf{Predicted}
    \\
		\hline\hline
		$r$ & $\gamma _{\mathrm{SBS}}$ & $\gamma _{\mathrm{GBS}}$
		\\
		\hline
		0.176 & 0.1867 & 0
		\\
		0.265 & 0.1923 & 0
		\\
		0.403 & 0.2054 & 0
		\\
		0.545 & 0.2250 & 0
		\\
		0.697 & 0.2531 & 0
		\\
		0.843 & 0.2873 & 0
		\\
        1.012 & 0.3355 & 0
        \\
        1.176 & 0.3904 & 0
        \\
		\hline\hline
	\end{tabular}
    \quad
    \begin{tabular}{lllll}
    \textbf{Measured}
    \\
		\hline\hline
		$r$ & $\gamma _{\mathrm{SBS}}$ & $\Delta \gamma _{\mathrm{SBS}}$ & $\gamma _{\mathrm{GBS}}$ & $\Delta \gamma _{\mathrm{GBS}}$
		\\
		\hline
		0.176 & 0.1788 & 0.0002 & 0.0011 & 0.0001
		\\
		0.265 & 0.1827 & 0.0002 & 0.0011 & 0.0001
		\\
		0.403 & 0.1932 & 0.0002 & 0.0012 & 0.0001
		\\
		0.545 & 0.2094 & 0.0002 & 0.0018 & 0.0001
		\\
		0.697 & 0.2331 & 0.0002 & 0.0026 & 0.0002
		\\
		0.843 & 0.2631 & 0.0002 & 0.0038 & 0.0002
		\\
        1.012 & 0.3062 & 0.0003 & 0.0064 & 0.0002
        \\
        1.176 & 0.3565 & 0.0004 & 0.0110 & 0.0002
        \\
		\hline\hline
	\end{tabular}
\end{table}

\subsection{Time dependence of measured data}

Here, we take a deeper look into the impact of phase noise on the measured minimum eigenvalues of our nonclassicality criterion. In particular, the timescale over which fluctuations are observed is explored. To this end, we plot the measured minimum eigenvalues obtained from the data for different averaging times.

\begin{figure}[!ht]
    \centering
    \includegraphics[width = \textwidth]{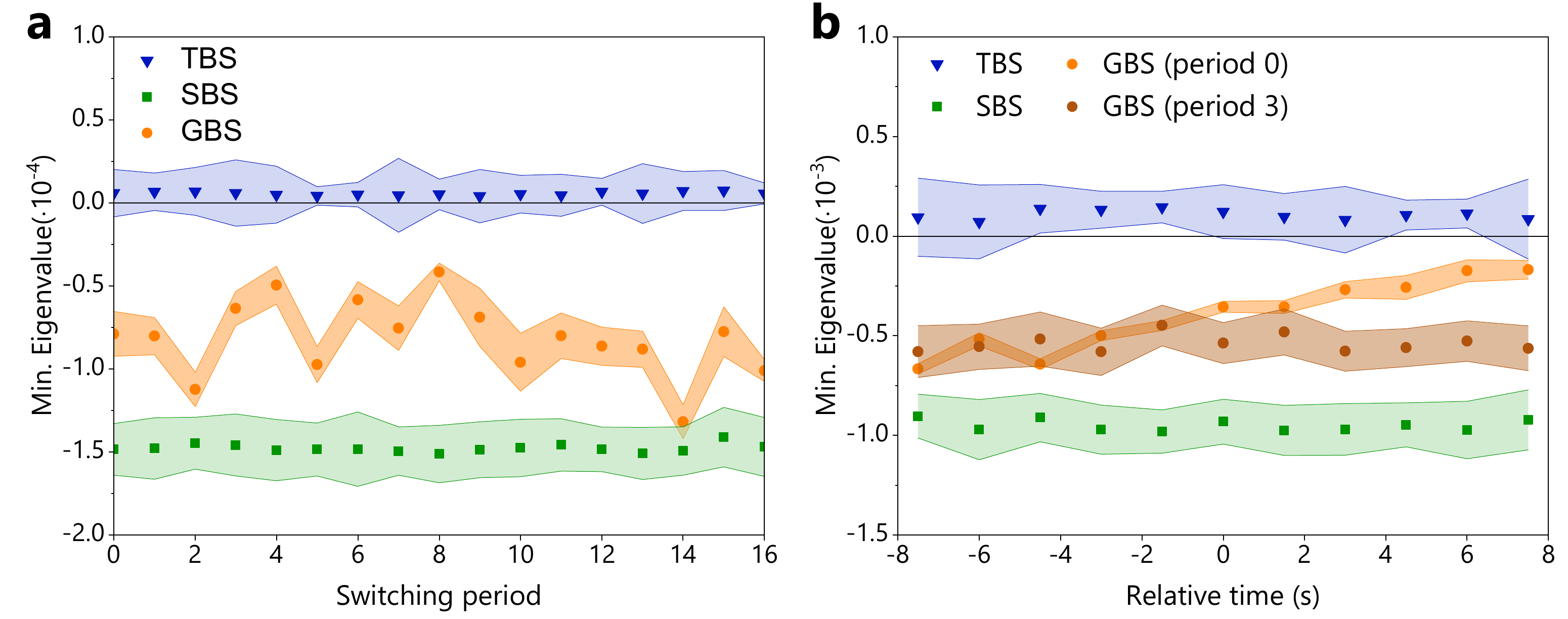}
    \caption{Characterization of Phase Drifts. \textbf{a,}  shows the minimum eigenvalues observed across the measurement run with a mean photon number of $\langle n \rangle = 0.031$. Each reported minimum eigenvalue represents 10\,s of integrated data from the center of the indicated switching period. \textbf{b,} shows the minimum eigenvalues observed within single 20\,s switching periods at a mean photon number of $\langle n \rangle = 0.172$. Here, each reported value has been integrated for a shorter time of 1\,s. The relative time indicates the center of the 1\,s integration time relative to the center of the corresponding switching period. GBS data for two different switching periods is presented. All uncertainties originate from counting errors.}
    \label{Fig:PhaseDrift}
\end{figure}

Figure \ref{Fig:PhaseDrift} (\textbf{a,} shows the minimum eigenvalues found for the $\langle n \rangle = 0.031$ measurement run when integrating data over 10\,s from the center of all 17 recorded switching periods (see Figure \ref{Fig:Pearson}). Therefore, we might expect some phase drift within the integration time and certainly expect the mean of the phases to drift between measurement points (See Figure \ref{Fig:PLSU_stability}), taken within subsequent switching periods, resulting in a temporal spacing between measurements of approximately 40\,s. The constant values obtained for the SBS and TBS data verify two things; firstly that these measures are insensitive to input phase drifts, and secondly that the other properties of the system, such as interference visibility and stability of the implemented unitary, are largely stable over the 12 minute measurement time. The minimum eigenvalues for the GBS data does not evolve smoothly from one data point to the next which indicates that the phase has drifted significantly between these measurement points.

In Figure \ref{Fig:PhaseDrift} (\textbf{b}) we investigate the evolution of the minimum eigenvalue over shorter timescales by integrating over 1\,s periods within a 20\,s switching period.  The relative time denotes the offset between the center of our 1\,s integration window and the center of the switching period. Once again, we see that TBS and SBS show no time dependence. The minimum eigenvalue obtained for the GBS configuration is shown for the data obtained from two different switching periods, illustrating that phase drifts (in the means of the input phases) are seen over this timescale in some measurements, but not in others. Furthermore, a smooth evolution of the minimum eigenvalue is seen in the case where this value drifts, which indicates that phase drifts in the system occur slower than our 1\,s averaging time.

\subsection{Summary of Results}

\label{App:summary}

In Table \ref{tab:MainResults} and Table \ref{tab:MainResultsTBS} we report the measured minimum eigenvalues and their uncertainties for all the complete set of measurement runs.

\begin{table}[!ht]
	\caption{Minimum eigenvalues obtained for GBS and SBS data. Numerical results for the complete set of GBS and SBS measurement runs across all investigated squeezing parameters $r$.
	Each result is followed by its absolute uncertainty obtained via propagation of counting errors.
	}\label{tab:MainResults}
	\begin{tabular}{lllllll}
		\textbf{GBS (min)}
		\\
		\hline\hline
		$r$  \quad  &  min. eigenvalue \quad & abs. uncert. \quad
		\\
		\hline
		0.176 & $-0.000144$ & $0.000036$
		\\
		0.265 & $-0.000309$ & $0.000021$
		\\
		0.403 & $-0.000666$ & $0.000026$
		\\
		0.545 & $-0.001194$ & $0.000189$
		\\
		0.697 & $-0.002118$ & $0.000148$
		\\
		0.843 & $-0.000191$ & $0.000359$
		\\
        1.012 & $-0.001030$ & $0.000283$
        \\
        1.176 & $-0.002212$ & $0.000462$
        \\
		\hline\hline
	\end{tabular}
    \quad
	\begin{tabular}{lllllll}
		\textbf{GBS (max)}\\
		\hline\hline
		$r$  \quad  &  min. eigenvalue \quad & abs. uncert. \quad
		\\
		\hline
		0.176 & $-0.000025$ & $0.000019$
		\\
		0.265 & $-0.000036$ & $0.000077$
		\\
		0.403 & $+0.000001$ & $0.000091$
		\\
		0.545 & $-0.000179$ & $0.000197$
		\\
		0.697 & $+0.000298$ & $0.000227$
		\\
		0.843 & $+0.003645$ & $0.000339$
		\\
        1.012 & $+0.007687$ & $0.000304$
        \\
        1.176 & $+0.017495$ & $0.000695$
        \\
		\hline\hline
	\end{tabular}
    \\[2ex]
	\begin{tabular}{lllllll}
		\textbf{SBS (min)}
		\\
		\hline\hline
		$r$  \quad  &  min. eigenvalue \quad & abs. uncert. \quad
		\\
		\hline
		0.176 & $-0.000194$ & $0.000053$
		\\
		0.265 & $-0.000404$ & $0.000084$
		\\
		0.403 & $-0.001024$ & $0.000111$
		\\
		0.545 & $-0.001666$ & $0.000160$
		\\
		0.697 & $-0.003749$ & $0.000225$
		\\
		0.843 & $-0.008245$ & $0.000290$
		\\
        1.012 & $-0.014337$ & $0.000430$
        \\
        1.176 & $-0.027048$ & $0.000708$
        \\
		\hline\hline
	\end{tabular}
    \quad
	\begin{tabular}{lllllll}
		\textbf{SBS (max)}\\
		\hline\hline
		$r$  \quad  &  min. eigenvalue \quad & abs. uncert. \quad
		\\
		\hline
		0.176 & $-0.000154$ & $0.000047$
		\\
		0.265 & $-0.000333$ & $0.000070$
		\\
		0.403 & $-0.000861$ & $0.000102$
		\\
		0.545 & $-0.001445$ & $0.000184$
		\\
		0.697 & $-0.003272$ & $0.000254$
		\\
		0.843 & $-0.005952$ & $0.000280$
		\\
        1.012 & $-0.012655$ & $0.000397$
        \\
        1.176 & $-0.022533$ & $0.000622$
        \\
		\hline\hline
	\end{tabular}
\end{table}

\clearpage

\begin{table}[!ht]
	\caption{Minimum eigenvalues obtained for TBS data. Numerical results for the complete set of TBS measurement runs across all investigated squeezing parameters $r$.
	Each result is followed by its absolute uncertainty obtained via propagation of counting errors.
	}\label{tab:MainResultsTBS}
	\begin{tabular}{lllllll}
		\textbf{TBS (min)}
		\\
		\hline\hline
		$r$  \quad  &  min. eigenvalue \quad & abs. uncert. \quad
		\\
		\hline
		0.176 & $-0.000014$ & $0.000047$
		\\
		0.265 & $-0.000012$ & $0.000068$
		\\
		0.403 & $+0.000062$ & $0.000028$
		\\
		0.545 & $+0.000319$ & $0.000298$
		\\
		0.697 & $+0.001023$ & $0.000423$
		\\
		0.843 & $+0.002591$ & $0.000581$
		\\
        1.012 & $+0.006680$ & $0.000823$
        \\
        1.176 & $+0.015123$ & $0.001150$
        \\
		\hline\hline
	\end{tabular}
    \quad
	\begin{tabular}{lllllll}
		\textbf{TBS (max)}\\
		\hline\hline
		$r$  \quad  &  min. eigenvalue \quad & abs. uncert. \quad
		\\
		\hline
		0.176 & $+0.000007$ & $0.000041$
		\\
		0.265 & $+0.000029$ & $0.000056$
		\\
		0.403 & $+0.000160$ & $0.000050$
		\\
		0.545 & $+0.000540$ & $0.000249$
		\\
		0.697 & $+0.001471$ & $0.000435$
		\\
		0.843 & $+0.003629$ & $0.000626$
		\\
        1.012 & $+0.008381$ & $0.000895$
        \\
        1.176 & $+0.018262$ & $0.001245$
        \\
		\hline\hline
	\end{tabular}
\end{table}



%

\end{document}